\documentclass[12pt,draftclsnofoot,onecolumn]{IEEEtran}
\usepackage[latin1]{inputenc}
\usepackage{graphicx}
\usepackage{color}
\usepackage{placeins}
\usepackage{float}
\usepackage{tabularx,colortbl}
\usepackage{amssymb}
\usepackage{amsthm}
\usepackage{cite}
\usepackage{amsmath}
\usepackage{caption2}

\usepackage{cases,subeqnarray}
\usepackage{bm,multirow,bigstrut}
\usepackage{textcomp}
\usepackage{latexsym,bm}
\usepackage{booktabs,changebar}
\usepackage{xcolor}
\usepackage{mathtools}
\usepackage{dsfont}
\usepackage{extarrows}

\usepackage{subfigure}

\usepackage{algorithm}
\usepackage{algpseudocode}

\theoremstyle{plain}
\newtheorem{thm}{Theorem}
\newtheorem{lemm}{Lemma}
\theoremstyle{plain}
\newtheorem{rem}{Remark}
\newtheorem{cor}{Corollary}

\IEEEoverridecommandlockouts

\begin{document}


\title{Cell-Free Massive MIMO-OFDM for High-Speed Train Communications}

\author{Jiakang Zheng,~\IEEEmembership{Student Member,~IEEE}, Jiayi~Zhang,~\IEEEmembership{Senior Member,~IEEE}, \mbox{Emil~Bj\"{o}rnson,~\IEEEmembership{Fellow,~IEEE}, Zhetao Li,~\IEEEmembership{Member,~IEEE}, and Bo Ai,~\IEEEmembership{Fellow,~IEEE}}
\thanks{J. Zheng and J. Zhang are with the School of Electronic and Information Engineering, Beijing Jiaotong University, Beijing 100044, P. R. China.}
\thanks{E. Bj\"{o}rnson is with the Department of Computer Science, KTH Royal Institute of Technology, Kista, Sweden.}
\thanks{Z. Li is with the Hunan International Scientific and Technological Cooperation Base of Intelligent network, and Key Laboratory of Intelligent Computing \& Information Processing of Ministry of Education, Xiangtan University, Hunan 411105, P. R. China.}
\thanks{B. Ai is with the State Key Laboratory of Rail Traffic Control and Safety, Beijing Jiaotong University, Beijing 100044, China.}
}

\maketitle
\vspace{-2cm}
\begin{abstract}
Cell-free (CF) massive multiple-input multiple-output (MIMO) systems show great potentials in low-mobility scenarios, due to cell boundary disappearance and strong macro diversity.
However, the great Doppler frequency offset (DFO) leads to serious inter-carrier interference in orthogonal frequency division multiplexing (OFDM) technology, which makes it difficult to provide high-quality transmissions for both high-speed train (HST) operation control systems and passengers.
In this paper, we focus on the performance of CF massive MIMO-OFDM systems with both fully centralized and local minimum mean square error (MMSE) combining in HST communications. Considering the local maximum ratio (MR) combining, the large-scale fading decoding (LSFD) cooperation and the practical effect of DFO on system performance, exact closed-form expressions for uplink spectral efficiency (SE) expressions are derived. We observe that cooperative MMSE combining achieves better SE performance than uncooperative MR combining.
In addition, HST communications with small cell and cellular massive MIMO-OFDM systems are compared in terms of SE. Numerical results reveal that the CF massive MIMO-OFDM system achieves a larger and more uniform SE than the other systems.
Finally, the train antenna centric (TA-centric) CF massive MIMO-OFDM system is designed for practical implementation in HST communications, and three power control schemes are adopted to optimize the propagation of TAs for reducing the impact of the DFO.
\end{abstract}
\vspace{-0.5cm}

\IEEEpeerreviewmaketitle

\newpage
\section{Introduction}

Cell-free (CF) massive multiple-input multiple-output (MIMO) has been proposed as a major leap of massive MIMO technology to circumvent the inter-cell interference, which is the main inherent limitation of dense cellular networks \cite{Ngo2017Cell}. CF massive MIMO systems consist of a large number of geographically distributed access points (APs) connected to a central processing unit (CPU), and coherently serve all user equipments (UEs) by spatial multiplexing on the same time-frequency resource \cite{chen2018channel}.
Therefore, CF massive MIMO systems can be deployed to guarantee good coverage and achieve almost uniform spectral efficiency (SE) without cells or cell edges \cite{zhang2020prospective}. The results in \cite{Ngo2017Cell} and \cite{8768014} revealed that the CF massive MIMO system achieves better performance than small cell and cellular massive MIMO systems in terms of 95\%-likely per-user SE.
A large amount of fundamental and important aspects of CF massive MIMO have been investigated in recent years. For example, \cite{bjornson2019making} studied both fully centralized and local processing in CF massive MIMO systems with minimum mean square error (MMSE) combining schemes. Although fully centralized processing can achieve maximal performance in CF massive MIMO systems, the huge processing complexity causes a great burden on the CPU \cite{9446982,9354156}. Then, authors in \cite{shaik2020mmse} proposed the sequential processing algorithm that achieves the same performance as the optimal centralized implementation, while reducing the fronthaul requirement. In addition, using the low complexity maximum ratio (MR) combining, novel expressions for CF massive MIMO systems with matched filter (MF) cooperation are derived in \cite{9453784}. Moreover, large-scale fading decoding (LSFD) cooperation is applied in CF massive MIMO systems to achieve two-fold performance gains than MF cooperation \cite{zheng2020efficient}.
The reason is that the interference statistics of the entire network are utilized to weigh the received signals at the APs and hence reduce interference.

Railway communications have attracted significant attention from both academia and industry due to the booming development of railways, especially high-speed train (HST) communications \cite{ai20205g}. Viaducts and tunnels are the two typical scenarios in wireless propagation environment for HST \cite{6469255}. Specifically, 86.5\% of railways are elevated in the Beijing-Shanghai HST \cite{7155729}. Therefore, there are few multi-path because of little scattering and reflection, and the line-of-sight (LoS) assumption is widely applicable in HST communications since the APs can be deployed to achieve that. For example, assuming small-scale fading is a constant value, \cite{7842154} studied the performance limits of uplink wireless delay-limited information transmission in HSTs under two trains encountering scenario, and \cite{7898862} analyzed the beamforming design principles for HST communications based on the location information.
However, in practice, the wireless propagation environment for HST is extremely diverse \cite{8847226}. In addition to the viaducts and tunnels, the train may travel in a variety of terrains such as train stations, mountains, forests and urban areas, which lead to abundant scattering components. Therefore, non-line-of-sight (NLoS) paths also need to be considered to accurately characterize the HST channel. For example, the scattering and reflection cannot be ignored in rich scattering terrains, and the small-scale fading can be well modeled with Nakagami-$m$ fading \cite{7505643}. Moreover, Rician fading based on $K$-factor can effectively describe the LoS and NLoS components of the wireless channel on HST communications \cite{6180104}. Therefore, Rician channel fading can capture some important insights of practical HST channels \cite{ai20205g}.

In addition, the orthogonal frequency-division multiplexing (OFDM) technique is applied to the long-term evolution for railway (LTE-R) for providing a seamless connection to existing ground cellular network \cite{7921554,7553613}. However, in high mobility scenarios, Doppler frequency offset (DFO), phase noise and timing offset can trigger the inter-carrier interference (ICI) and seriously decrease the performance of systems \cite{6469255,9205984,6189004}. Moreover, fast signal processing in light of high mobility and the frequency handover between adjacent base stations (BSs) are two of the main challenges in HST system design \cite{6180090}. One of the promising solutions to tackle these challenges is to utilize a distributed massive MIMO architecture \cite{7355284,9598918}.
Simulation results in \cite{6712160} showed that, for ground-train communications, the distributed antenna system structure can reduce the unnecessary handovers, as well as reduce the handover failure and link outage probabilities. In fact, for distributed antenna systems, enhanced simplified fast handover schemes are still needed to deliver ultra-reliable and low-latency HST communications  \cite{9272878}.
However, as a new framework of distributed massive MIMO architecture, CF massive MIMO systems completely eliminate the cell edge, which is very suitable for HST communications \cite{ai20205g}.
More importantly, the authors in \cite{9416909} pointed out that, in both static and mobile scenarios, the CF massive MIMO system performs better than traditional cellular-based networks.
In addition, the CF massive MIMO-OFDM system was presented and analyzed in \cite{9446530}, which provides a baseline for further studying. Besides, authors in \cite{gao2020uplink} and \cite{9646498} investigated the crowded CF massive MIMO-OFDM system with spatially and frequently correlated channels and hybrid beamformers of CF massive MIMO-OFDM systems over millimeter-wave channels, respectively. However, the important ICI problem is neglected for both of them. Therefore, the fundamental limits of HST communications with CF massive MIMO-OFDM systems are still an open question.

Although with promising benefits, HST communications with CF massive MIMO systems are also faced with a lot of problems in practical deployment \cite{ai20205g}. For instance, the total computational complexity and fronthaul requirements approach infinity as the number of AP increases infinitely with the length of railway line \cite{8598980,8901451}. Hence, the optimal AP cooperation and power control algorithms are not feasible for practical implementation. Utilizing the user-centric clusters method, a new framework named scalable CF massive MIMO systems was proposed in \cite{bjornson2020scalable} to achieve the benefits of CF operation in a practically feasible way, with computational complexity and fronthaul requirements that are scalable to large networks. In addition, the paper \cite{9174860} provided a feasible solution for structured massive access in CF massive MIMO systems to achieve higher SE to more UEs. Meanwhile, efficient and low complexity scalable power control algorithms were provided in \cite{9399102} to improve performance gain in a very large service area. However, the linear structure of the high-speed railway is different from the large square service area considered in the previous work. That is, the practical application of CF massive MIMO systems in HST communications needs to be carefully designed.

Motivated by the aforementioned observation, we investigate the performance of the CF massive MIMO-OFDM system in HST communications, where high mobility may destroy the orthogonality of subcarriers to cause serious ICI. Fully centralized and local processing with both MR and MMSE combining schemes are considered in the uplink data transmission. The performance of HST communications with conventional small cell and cellular massive MIMO-OFDM systems are analyzed for comparison. Then, we design the train antenna centric (TA-centric) CF massive MIMO system in a linear structured HST scenario for practical implementation. The specific contributions of the work are listed as follows:

\begin{itemize}
  \item We first derive closed-form expressions for the SE of the CF massive MIMO-OFDM system with ICI effect in HST communications. Our results show that effective interference cancellation schemes like MMSE combining and LSFD cooperation are very necessary for HST communications. In addition, the CF massive MIMO system achieves higher and more uniform SE performance than small cell and cellular systems in HST communications.
  \item We find that LoS components are helpful to improve SE performance, but it is more vulnerable to the effect of DFO. Increasing the number of APs significantly improves the average SE and larger distance between the rail track and APs is preferred for the optimal operating point of SE. In addition, decreasing the number of TAs and increasing the number of antennas per AP both are available ways to reduce the influence of DFO.
  \item We design the TA-centric CF massive MIMO system in a linear structured HST scenario for practical implementation. It is found that the average SE of the worst TA increases with the increase of the number of access APs. However, the benefits of the practical fractional power control are gradually weakened as the number of access APs increases.
\end{itemize}

Note that the conference version of this paper \cite{zheng2022uplink} investigated the uplink CF massive MIMO-OFDM systems in HST communications with the LoS channel.
The rest of the paper is organized as follows. In Section \ref{se:model}, we describe the HST communication model incorporating the combined effects of channel estimation error, spatial correlation and ICI. Next, Section \ref{se:performance} presents the achievable uplink SE of HST with ICI effect for all CF, SC and cellular systems. Then, Section \ref{se:Scalable} investigates the TA-centric CF massive MIMO-OFDM system in HST communications, as well as the fractional power control used to improve the SE performance of the worst TA. We provide numerical results and discussions in Section \ref{se:NR}. Finally, Section \ref{se:CON} gives a brief summary and provides suggestions for future work.

\textit{Notation:} We use boldface lowercase letters $\mathbf{x}$ and boldface uppercase letters $\mathbf{X}$ to represent column vectors and matrices, respectively.
Superscripts $x^\mathrm{*}$, $\mathbf{x}^\mathrm{T}$ and $\mathbf{x}^\mathrm{H}$ are used to denote conjugate, transpose and conjugate transpose, respectively.
The absolute value, the Euclidean norm, the expectation operator, the trace operator and the definitions are denoted by $\left|  \cdot  \right|$, $\left\|  \cdot  \right\|$, $\mathbb{E}\left\{  \cdot  \right\}$, ${\text{tr}}\left(  \cdot  \right)$, and $\triangleq$, respectively.
${{\mathbf{I}}_N}$ is the $N\times N$ identity matrix, and ${\log _2}\left(  \cdot  \right)$ denotes the logarithm with base 2.
Finally, $x \sim \mathcal{C}\mathcal{N}\left( {0,{\sigma^2}} \right)$ represents a circularly symmetric complex Gaussian random variable $x$ with variance $\sigma^2$.
\section{System Model}\label{se:model}

We consider a HST communication with CF massive MIMO-OFDM system consisting of $K$ distributed TAs on the top of different cars of a HST (length of the entire train is $d_\text{hst}$) and $L$ APs (uniform linear array) along one side of the rail track, where each AP is equipped with $N$ antennas.\footnote{Multiple-antennas APs are good for designing high-performance beamforming algorithms to provide linear coverage \cite{7553613}.
In addition, the derived results of considered one-antenna TAs can be seen as a lower bound of the performance of multi-antennas.} The APs are connected to a CPU via fronthaul links. We assume that all $L$ APs serve all $K$ TAs on the same time-frequency resource and all antennas at the same height above the ground. The propagation loss increases with the distance, which causes that only nearby APs or the closest BS contribute. Therefore, we here only consider a section of the railway with length $d_\text{rai} = 5 d_\text{hst}$ for analysis. How TAs can be served on the entire track with thousands of APs is investigated in Section \ref{se:Scalable}.

\begin{figure}[t]
\centering
\includegraphics[scale=0.9]{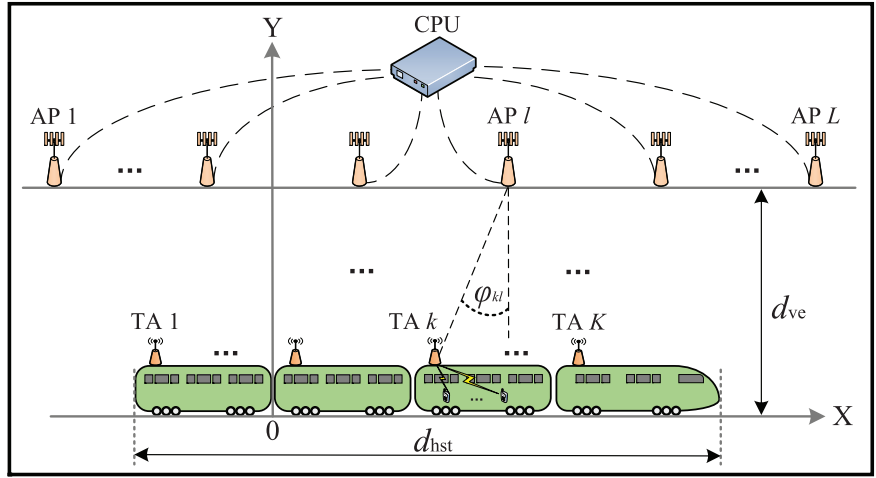}
\caption{HST communications with a cooperative CF massive MIMO-OFDM system.} \vspace{-4mm}
\label{system_model}
\end{figure}

The track of HST is mostly straight, and will not turn in a short distance, so we assume our considered section of railway is a straight line\footnote{Straight and curved route shapes are typical in HST scenarios, but the range of train speed is 0 km/h-120 km/h in the curved route scenario \cite{8168415}. Therefore, the assumption of straight line complies with reality in our considered high-mobility scenarios.}. As illustrated in Fig.~\ref{system_model}, a 2D plane coordinate system is used to determine the locations of $L$ APs and $K$ TAs. We assume the coordinates of APs are ${{\mathbf{q}}_l} \triangleq \left[ {{a_l},{d_\text{ve}}} \right],l = 1, \ldots ,L$, where ${a_l}$ is the horizontal coordinate of AP $l$, and $d_\text{ve}$ is the distance between the rail track and APs. In addition, the coordinates of TAs are ${{\mathbf{q}}_k} \triangleq \left[{\left({a_k} + d_\text{tr} \right),0} \right],k = 1, \ldots ,K$, where ${a_k}$ is the initial abscissa position of TA $k$, and ${d_{{\text{tr}}}} \in \left[ { - \infty , + \infty } \right]$ is the distance traveled by the train, which indicates the position of the HST. Note that $d_\text{tr}$ can be positive or negative, which represent the forward and backward of the HST, respectively. Moreover, we can obtain the straight line distance between AP $l$ and TA $k$ as $d_{kl} = \left\| {{{\mathbf{q}}_l} - {{\mathbf{q}}_k}} \right\|$, and abscissa difference between them is $d_{kl}^{{\text{ho}}} = {a_l} - {a_k}$. Furthermore, the angle of arrival (AOA) from TA $k$ to AP $l$ is denoted by $\varphi_{kl}$, and then the sine value of it can be given by $\sin \left( {{\varphi _{kl}}} \right) = {{d_{kl}^{{\text{ho}}}}}/{{{d_{kl}}}}$.
Besides, in Fig.~\ref{system_model}, four different levels of cooperation can be achieved based on different signal processing methods at AP and CPU \cite{bjornson2019making}.

\subsection{Propagation Model}

\begin{figure}[t]
\centering
\includegraphics[scale=1.5]{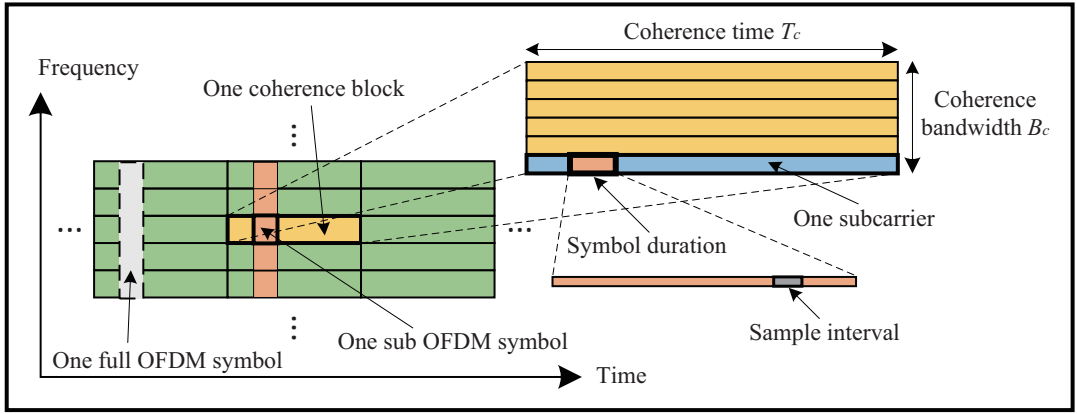}
\caption{OFDM transmission with ICI between adjacent subcarriers.} \vspace{-4mm}
\label{block}
\end{figure}

Considering the MIMO-OFDM channel in the HST scenario as illustrated in Fig. \ref{block}, there exists ICI between adjacent subcarriers. The bandwidth is $B$, the number of total subcarriers is $M_\text{total}$, and the duration of one OFDM symbol is $T_{s}$.
The coherence bandwidth and the coherence time are $B_c$ and $T_c$, respectively. Each coherence block is flat-fading, and the random channel responses in one coherence block are statistically identical to the ones in any other coherence block, as well as the channel realizations are independent between any pair of blocks. Hence, the performance analysis is therefore carried out by studying a single statistically representative coherence block.
\begin{rem}
We assume large communication bandwidths are used so that we can transmit sufficiently long codewords to approach the ergodic capacity also in short time segments by coding over the frequency domain.
\end{rem}
The HST channel contains one LoS path and a large number of NLoS paths.
We assume that there is a large number of statistically independent reflected and scattered paths with random amplitudes in the delay window corresponding to a single channel filter tap so that a statistical model can be used \cite{tse2005fundamentals}.
Therefore, the channel gain between AP $l$ and TA $k$ can be modeled by Rician fading as ${{\mathbf{g}}_{kl}} \sim \mathcal{C}\mathcal{N}\left( {{{{\mathbf{\bar h}}}_{kl}},{{\mathbf{R}}_{kl}}} \right)$, where ${{{\mathbf{\bar h}}}_{kl}} \in {\mathbb{C}^N}$ is the LoS component with a known magnitude and uniform phase, and ${{\mathbf{R}}_{kl}} \in {\mathbb{C}^{N \times N}}$ is the positive semi-definite covariance matrix describing the spatial correlation of the NLoS components.
It is worth noting that channel gain also can be expressed as ${{\mathbf{g}}_{kl}} = {{{\mathbf{\bar h}}}_{kl}} + {{\mathbf{h}}_{kl}}$, where ${{\mathbf{h}}_{kl}} \sim \mathcal{C}\mathcal{N}\left( {{\mathbf{0}},{{\mathbf{R}}_{kl}}} \right)$ is a complex Gaussian distributed component. Then, as the HST travels along the railway, the Rician channel fading with DFO effect can be modeled as following:

High speed movement in HST scenario will lead to large DFO and shorter coherence blocks.
The DFO effect may create a correlation between the fading realization at adjacent subcarriers that is also gradually decaying.
Therefore, as illustrated in Fig.~\ref{block}, we can use a full OFDM transmission in real systems, but only study one block of it for approximate analysis with small errors.
Considering one coherence block with $M$ subcarriers, where ${{\mathbf{g}}_{il}}\left[ 1 \right] =  \ldots  = {{\mathbf{g}}_{il}}\left[ M \right] = {{\mathbf{g}}_{il}}$ at different subcarriers, but there are ICI effect among them.
The information bit stream is multiplexed into $M$ symbol streams, each with symbol duration $T_s$, modulating a set of $M$ subcarriers that are spaced by $1/T_{s}$.
\begin{lemm}
We assume different signals are sent at different subcarriers.
Then, the total received signal from all TAs at AP $l$ can be expressed in the frequency domain as
\begin{align}\label{y_l}
{{\mathbf{y}}_l}\left[ s \right] = \sum\limits_{i = 1}^K {\sum\limits_{m = 1}^M {\sqrt {{p_i}} } \left( {{I_{il}}\left[ {m - s} \right]{{{\mathbf{\bar h}}}_{il}} + {I_{{D}}}\left[ {m - s} \right]{{\mathbf{h}}_{il}}} \right){x_i}} \left[ m \right] + {{\mathbf{w}}_l}\left[ s \right],
\end{align}
with
\begin{align}
\label{I_kl}{I_{il}}\left[ {m - s} \right] &= \frac{{\sin \left( {\pi \left( {m - s+ {\varepsilon _{il}}} \right)} \right)}}{{M\sin \left( {\frac{\pi }{M}\left( {m - s+ {\varepsilon _{il}}} \right)} \right)}} \exp \left( {j\pi \left( {1 - \frac{1}{{\;M}}} \right)\left( {m - s+ {\varepsilon _{il}}} \right)} \right), \\
\label{NLOSD}{I_{{D}}}\left[ {m - s} \right] &= \left\{ {\begin{array}{*{20}{c}}
  {1\;\;\;\;\;\;\;\;\;\;\;\;\;\;,m = s}, \\
  {\frac{{{{\left( { - 1} \right)}^{m - s}}\omega}}{{\sqrt 2 \left( {m - s} \right)}}\;\;\;,m \ne s},
\end{array}} \right.
\end{align}
where ${\varepsilon _{il}} = \omega\sin \left( {{\varphi _{il}}} \right) = \frac{{fvT_s}}{c}\sin \left( {{\varphi _{il}}} \right)$ is the normalized DFO between the $l$th receive AP and the $i$th transmit TA. Besides, $v$ and $c$ are the velocity of train and light, respectively. Furthermore, $f$ and $T_s$ denote the carrier frequency and the time domain duration of one OFDM symbol.
\end{lemm}
\begin{IEEEproof}
Please refer to Appendix A.
\end{IEEEproof}
It is found that the variance of ICI coefficients of adjacent subcarriers are less than $10^{-2}$ from the fourth. Therefore, $M\geqslant8$ must be guaranteed to obtain a good approximation.

\subsection{Uplink Channel Estimation}

We assume that $\tau_p $ ($\tau_p=K$)\footnote{We do not consider the pilot contamination in HST communications, because the pilots are enough to be allocated to a limited number of TAs on one train. In addition, the use of non-orthogonal pilots does not affect our analysis of DFO but will bring more complex formulas.} mutually orthogonal $\tau_p$ length pilot signals $\bm{\phi}_1, \ldots , \bm{\phi}_{\tau_p}$ with ${\left\| {\bm{\phi}_i } \right\|^2} = {\tau _p}$ are used for channel estimation\footnote{We assume that each pilot sample is staggered in time to ensure that the orthogonality is not destroyed by ICI. Therefore, the DFO has no effect on the channel estimation.}. The received signal of AP $l$ is ${{\mathbf{Z}}_l} = \sum\limits_{i = 1}^K {\sqrt {{p_i}} } {{\mathbf{g}}_{il}}\bm{\phi}_i^{\text{T}} + {{\mathbf{N}}_l}$, where $p_i \geqslant 0$ is the transmit power of TA $i$, and ${{\mathbf{N}}_l} \in {\mathbb{C}^{N \times {\tau _p}}}$ is the receiver noise with independent $\mathcal{C}\mathcal{N}\left( {0,{\sigma ^2}} \right)$ entries, where $\sigma^2$ is the noise power. To estimate $\mathbf{g}_{kl} $, the AP first correlates the received signal with the associated normalized pilot signal ${\bm{\phi} _k}/\sqrt {{\tau _p}}$ to obtain
\begin{align}
  {{\mathbf{z}}_{kl}} = \frac{1}{{\sqrt {{\tau _p}} }}{{\mathbf{Z}}_l}\bm{\phi} _k^* = \frac{1}{{\sqrt {{\tau _p}} }}\left( {\sum\limits_{i = 1}^K {\sqrt {{p_i}} } {{\mathbf{g}}_{il}}\bm{\phi} _i^{\text{T}} + {{\mathbf{N}}_l }} \right)\bm{\phi}_k^*  = \sqrt {{p_k}{\tau _p}} \left( {{{{\mathbf{\bar h}}}_{kl}} + {{\mathbf{h}}_{kl}}} \right) + {{\mathbf{n}}_{kl}}.
\end{align}
Using standard results from estimation theory \cite{bjornson2017massive}, the phase-aware\footnote{In view of the highly deterministic propagation environment of the HST scenarios, where the speed, direction and position of the train can be detected by a track circuit \cite{7553613}, we assume the phase-shift of LoS link is known by APs and CPU.} MMSE estimate of $\mathbf{g}_{kl}$ is
\begin{align}
  {{{\mathbf{\hat g}}}_{kl}} = {{{\mathbf{\bar h}}}_{kl}} + {{{\mathbf{\hat h}}}_{kl}} = {{{\mathbf{\bar h}}}_{kl}} + \sqrt {{p_k}{\tau _p}} {{\mathbf{R}}_{kl}}{{\mathbf{\Psi }}_{kl}}\left( {{{\mathbf{z}}_{kl}} - {{{\mathbf{\bar z}}}_{kl}}} \right) ,
\end{align}
where ${{{\mathbf{\bar z}}}_{kl}} = \sqrt {{p_k}{\tau _p}} {{{\mathbf{\bar h}}}_{kl}}$, and ${{\mathbf{\Psi }}_{kl}} = {\left( {{p_k}{\tau _p}{{\mathbf{R}}_{kl}} + {\sigma ^2}{{\mathbf{I}}_N}} \right)^{ - 1}}$.
Moreover, the estimate ${{{\mathbf{\hat g}}}_{kl}}$ and the estimation error ${{{\mathbf{\tilde g}}}_{kl}} = {{\mathbf{g}}_{kl}} - {{{\mathbf{\hat g}}}_{kl}}$ are respectively distributed as $\mathcal{C}\mathcal{N}\left( {{{{\mathbf{\bar h}}}_{kl}},{{\mathbf{Q}}_{kl}}} \right)$ and $\mathcal{C}\mathcal{N}\left( {{\mathbf{0}},{{\mathbf{C}}_{kl}}} \right)$, where $ {{\mathbf{Q}}_{kl}} = {p_k}{\tau _p}{{\mathbf{R}}_{kl}}{{\mathbf{\Psi }}_{kl}}{{\mathbf{R}}_{kl}} $, and ${{\mathbf{C}}_{kl}} = {{\mathbf{R}}_{kl}} - {{\mathbf{Q}}_{kl}}$.

\section{Performance Analysis}\label{se:performance}

In this section, we study the uplink SE of CF massive MIMO-OFDM systems in HST communications. Meanwhile, small cell and cellular massive MIMO-OFDM systems are analyzed for comparison. We also derive closed-form SE expressions for all considered systems. Note that we assume different TAs and subcarriers send different signals in data transmission, which leads to both inter-user interference and ICI. In the following analysis, one time interval is considered, but the obtained conclusions are valid for all time intervals by updating settings.

\subsection{CF Massive MIMO-OFDM With Fully Centralized Processing}

The most advanced level of CF massive MIMO operation is when the $L$ APs send their received pilot signals and received data signals to the CPU, which takes care of the channel estimation and data signal detection. In other words, the APs act as relays that forward all signals to the CPU.
Then, the received signal available at the CPU is expressed as
\begin{align}
\underbrace {\left[\! {\begin{array}{*{20}{c}}
  {{{\mathbf{y}}_1}\left[ s \right]} \\
   \vdots  \\
  {{{\mathbf{y}}_L}\left[ s \right]}
\end{array}} \!\right]}_{{\mathbf{y}}\left[ s \right]} \!=\! \sum\limits_{i = 1}^K {\sum\limits_{m = 1}^M {\sqrt {{p_i}} } \left( {{\mathbf{I}_i}\left[ {m - s} \right]\underbrace {\left[\! {\begin{array}{*{20}{c}}
  {{{{\mathbf{\bar h}}}_{i1}}} \\
   \vdots  \\
  {{{{\mathbf{\bar h}}}_{iL}}}
\end{array}} \!\right]}_{{{{\mathbf{\bar h}}}_i}} + {I_{{D}}}\left[ {m - s} \right]\underbrace {\left[\! {\begin{array}{*{20}{c}}
  {{{\mathbf{h}}_{i1}}} \\
   \vdots  \\
  {{{\mathbf{h}}_{iL}}}
\end{array}} \!\right]}_{{{\mathbf{h}}_i}}} \right){x_i}} \left[ m \right] + \underbrace {\left[\! {\begin{array}{*{20}{c}}
  {{{\mathbf{w}}_1}\left[ s \right]} \\
   \vdots  \\
  {{{\mathbf{w}}_L}\left[ s \right]}
\end{array}} \!\right]}_{{\mathbf{w}}\left[ s \right]}
\end{align}
or, in a more compact form, as
\begin{align}
{\mathbf{y}}\left[ s \right] = \sum\limits_{i = 1}^K {\sum\limits_{m = 1}^M {\sqrt {{p_i}} } \left( {{\mathbf{I}_i}\left[ {m - s} \right]{{{\mathbf{\bar h}}}_i} + {I_{{D}}}\left[ {m - s} \right]{{\mathbf{h}}_i}} \right){x_i}} \left[ m \right] + {\mathbf{w}}\left[ s \right],
\end{align}
where the collective channel is distributed as ${{\mathbf{h}}_k} \sim \mathcal{C}\mathcal{N}\left( {{\mathbf{0}},{{\mathbf{R}}_k}} \right)$ with ${{\mathbf{R}}_k} = {\text{diag}}\left( {{{\mathbf{R}}_{k1}}, \ldots ,{{\mathbf{R}}_{kL}}} \right)$ is the block-diagonal spatial correlation matrix. Moreover, we have
\begin{align}
{\mathbf{I}_i}\left[ {m - s} \right] \triangleq {\text{diag}}\left( {{I_{i1}}\left[ {m - s} \right]{{\mathbf{I}}_N}, \ldots ,{I_{iL}}\left[ {m - s} \right]{{\mathbf{I}}_N}} \right) .
\end{align}
The CPU can compute all the MMSE channel estimates using the received pilot signals and channel statistics obtained from the APs. For TA $k$, the CPU can then form the collective channel estimate ${{{\mathbf{\hat g}}}_k} \sim \mathcal{C}\mathcal{N}\left( {{{{\mathbf{\bar h}}}_k},{{\mathbf{Q}}_k}} \right)$, where ${{\mathbf{Q}}_k} = {\text{diag}}\left( {{{\mathbf{Q}}_{k1}}, \ldots ,{{\mathbf{Q}}_{kL}}} \right)$.
The estimation error is ${{{\mathbf{\tilde g}}}_k}{\text{ = }}{{\mathbf{g}}_k} - {{{\mathbf{\hat g}}}_k} \sim \mathcal{C}\mathcal{N}\left( {{\mathbf{0}},{{\mathbf{C}}_k}} \right)$ with ${{\mathbf{C}}_k} = {{\mathbf{R}}_k} - {{\mathbf{Q}}_k} = {\text{diag}}\left( {{{\mathbf{C}}_{k1}}, \ldots ,{{\mathbf{C}}_{kL}}} \right)$.
Next, the CPU selects an arbitrary receive combining vector for TA $k$ based on all the collective channel estimates.
The received signal at the CPU is
\begin{align}
  {y_k}\left[ s \right] &= {\mathbf{ v}}_k^{\text{H}}\left[ s \right]{\mathbf{y}}\left[ s \right] = \sqrt {{p_k}} {\mathbf{v}}_k^{\text{H}}\left[ s \right]\left( {{I_k}\left[ 0 \right]{{{\mathbf{\bar h}}}_k} + {I_{{D}}}\left[ 0 \right]{{{\mathbf{\hat h}}}_k}} \right){x_k}\left[ s \right] + \sqrt {{p_k}} {\mathbf{v}}_k^{\text{H}}\left[ s \right]{I_{{D}}}\left[ 0 \right]{{{\mathbf{\tilde h}}}_k}{x_k}\left[ s \right] \notag \\
   &+ \sum\limits_{m \ne s}^M {\sqrt {{p_k}} } {\mathbf{v}}_k^{\text{H}}\left[ s \right]\left( {{I_k}\left[ {m - s} \right]{{{\mathbf{\bar h}}}_k} + {I_{{D}}}\left[ {m - s} \right]{{\mathbf{h}}_k}} \right){x_k}\left[ m \right] \notag \\
   &+ \sum\limits_{i \ne k}^K {\sum\limits_{m = 1}^M {\sqrt {{p_i}} } {\mathbf{v}}_k^{\text{H}}\left[ s \right]\left( {{I_i}\left[ {m - s} \right]{{{\mathbf{\bar h}}}_i} + {I_{{D}}}\left[ {m - s} \right]{{\mathbf{h}}_i}} \right){x_i}} \left[ m \right] + {\mathbf{v}}_k^{\text{H}}\left[ s \right]{\mathbf{w}}\left[ s \right] .
\end{align}
\begin{lemm}\label{thm1}
If the MMSE estimator is used, an achievable uplink SE of TA $k$ is ${\mathrm{S}}{{\mathrm{E}}_k^\mathrm{fc}}\left[ s \right] = \mathbb{E}\left\{ {{{\log }_2}\left( {1 + \mathrm{SINR}_k^\mathrm{fc}\left[ s \right]} \right)} \right\}$ with $\mathrm{SINR}_k^\mathrm{fc}\left[ s \right]$ is given as
\begin{align}\label{SINR_f}
 \mathrm{SINR}_k^\mathrm{fc}\left[ s \right] &= {p_k}{\left| {{\mathbf{v}}_k^{\mathrm{H}}\left[ s \right]{{\mathbf{f}}_k}\left[ 0 \right]} \right|^2}\left( {{p_k}\sum\limits_{m \ne s}^M {{{\left| {{\mathbf{v}}_k^{\mathrm{H}}\left[ s \right]{{\mathbf{f}}_k}\left[ {m - s} \right]} \right|}^2}}  + \sum\limits_{i \ne k}^K {{p_i}\sum\limits_{m = 1}^M  {{\left| {{\mathbf{v}}_k^{\mathrm{H}}\left[ s \right]{{\mathbf{f}}_i}\left[ {m - s} \right]} \right|}^2}} } \right. \notag \\
  &{\left. { + {\mathbf{v}}_k^{\mathrm{H}}\left[ s \right]\left( {\sum\limits_{i = 1}^K {{p_i}\sum\limits_{m = 1}^M  I_{\text{D}}^2\left[ {m - s} \right]{{\mathbf{C}}_i} + {\sigma ^2}{{\mathbf{I}}_{LN}}} } \right){{\mathbf{v}}_k}\left[ s \right]} \right)^{ - 1}} ,
\end{align}
where
\begin{align}
{{\mathbf{f}}_i}\left[ {m - s} \right] \triangleq {I_i}\left[ {m - s} \right]{{{\mathbf{\bar h}}}_i} + {I_{\text{D}}}\left[ {m - s} \right]{{{\mathbf{\hat h}}}_i} .
\end{align}
\end{lemm}
\begin{IEEEproof}
The proof follows the same steps as the proof of \cite[Theorem 4.1]{bjornson2017massive} for Cellular mMIMO and is therefore omitted.
\end{IEEEproof}

\begin{cor}\label{cor1}
The instantaneous SINR for TA $k$ is maximized by the MMSE combining vector
\begin{align}
{{\mathbf{v}}_k\left[s\right]} = {p_k}{\left( {\sum\limits_{i = 1}^K {{p_i}\sum\limits_{m = 1}^M {\left( {{{\mathbf{f}}_i}\left[ {m - s} \right]{\mathbf{f}}_i^{\mathrm{H}}\left[ {m - s} \right] + I_{\text{D}}^2\left[ {m - s} \right]{{\mathbf{C}}_i}} \right) + {\sigma ^2}{{\mathbf{I}}_{LN}}} } } \right)^{ - 1}}{{\mathbf{f}}_k}\left[ 0 \right],
\end{align}
which leads to the maximum performance ${\mathrm{S}}{{\mathrm{E}}_k^\mathrm{fc,mmse}}\left[ s \right] = \mathbb{E}\left\{ {{{\log }_2}\left( {1 + \mathrm{SINR}_k^\mathrm{fc,mmse}\left[ s \right]} \right)} \right\}$, with \\ $\mathrm{SINR}_k^\mathrm{fc,mmse}\left[ s \right]$ is given as
\begin{align}
 {p_k}{\mathbf{f}}_k^{\mathrm{H}}\left[ 0 \right]\left( {{p_k}\sum\limits_{m \ne s}^M {{{\mathbf{f}}_k}\left[ {m - s} \right]{\mathbf{f}}_k^{\mathrm{H}}\left[ {m - s} \right] + \sum\limits_{i \ne k}^K {{p_i}\sum\limits_{m = 1}^M {{{\mathbf{f}}_i}\left[ {m - s} \right]{\mathbf{f}}_i^{\mathrm{H}}\left[ {m - s} \right]} } } } \right. \notag \\
 {\left. { + \sum\limits_{i = 1}^K {{p_i}\sum\limits_{m = 1}^M {I_{\text{D}}^2\left[ {m - s} \right]{{\mathbf{C}}_i} + {\sigma ^2}{{\mathbf{I}}_{LN}}} } } \right)^{ - 1}}{{\mathbf{f}}_k}\left[ 0 \right] .
\end{align}
\end{cor}
\begin{IEEEproof}
It follows from \cite[Lemma B.10]{bjornson2017massive} since \eqref{SINR_f} is a generalized Rayleigh quotient with respect to $\mathbf{v}_k\left[s\right]$.
\end{IEEEproof}
\begin{rem}
To reduce the fronthaul capacity of CF massive MIMO with fully centralized processing, a novel uplink sequential processing algorithm is developed in \cite{shaik2020mmse}. It achieves the same optimal performance as centralized MMSE but with lower fronthaul requirements.
\end{rem}

\subsection{CF Massive MIMO-OFDM With Local Processing}

Each AP can preprocess its signal by computing local estimates of the data that are then passed to the CPU for final decoding. Let ${{\mathbf{v}}_{kl}\left[s\right]} \in {\mathbb{C}^N}$ be the local combining vector that AP $l$ selects for TA $k$ at the $s$th subcarrier. Then, based on the received signal \eqref{y_l}, its local estimate ${{\overset{\lower0.5em\hbox{$\smash{\scriptscriptstyle\smile}$}}{y} }_{kl}}\left[ s \right]  =  {\mathbf{v}}_{kl}^{\text{H}}\left[ s \right]{{\mathbf{y}}_l}\left[ s \right] $ is obtained as
\begin{align}
{{\overset{\lower0.5em\hbox{$\smash{\scriptscriptstyle\smile}$}}{y} }_{kl}}\left[ s \right] = \sum\limits_{i = 1}^K {\sum\limits_{m = 1}^M {\sqrt {{p_i}} } {\mathbf{v}}_{kl}^{\text{H}}\left[ s \right]\left( {{I_{il}}\left[ {m - s} \right]{{{\mathbf{\bar h}}}_{il}} + {I_{D}}\left[ {m - s} \right]{{\mathbf{h}}_{il}}} \right){x_i}} \left[ m \right] + {\mathbf{v}}_{kl}^{\text{H}}\left[ s \right]{{\mathbf{w}}_l}\left[ s \right] .
\end{align}
The combining vector that minimizes the MSE, is
\begin{align}\label{v_kl}
{{\mathbf{v}}_{kl}}\left[ s \right] = {p_k}{\left( {\sum\limits_{i = 1}^K {{p_i}\sum\limits_{m = 1}^M {\left( {{\mathbf{f}_{il}}\left[ {m - s} \right]{{{\mathbf{f}_{il}^{\text{H}}}\left[ {m - s} \right]}} + I_{D}^2\left[ {m - s} \right]{{\mathbf{C}}_{il}}} \right) + {\sigma ^2}{{\mathbf{I}}_N}} } } \right)^{ - 1}}{\mathbf{f}_{kl}}\left[ {0} \right] ,
\end{align}
where
\begin{align}
{\mathbf{f}_{il}}\left[ {m - s} \right] \triangleq {{I_{il}}\left[ {m - s} \right]{{{\mathbf{\bar h}}}_{il}} + {I_{{D}}}\left[ {m - s} \right]{{{\mathbf{\hat h}}}_{il}}} .
\end{align}
The local estimates ${{\overset{\lower0.5em\hbox{$\smash{\scriptscriptstyle\smile}$}}{y} }_{kl}}\left[ s \right]$ are then sent to the CPU where they are linearly combined using the LSFD weights ${a_{kl}\left[ s \right]}$ to obtain ${{\hat y}_k}\left[ s \right] = \sum\limits_{l = 1}^L {a_{kl}^*\left[ s \right]{{\overset{\lower0.5em\hbox{$\smash{\scriptscriptstyle\smile}$}}{y} }_{kl}}\left[ s \right]} $, which is eventually used to decode $x_k\left[ s \right]$. We have that
\begin{align}
{{\hat y}_k}\left[ s \right] = \sum\limits_{i = 1}^K {\sum\limits_{m = 1}^M {\sqrt {{p_i}} } {\mathbf{a}}_k^{\text{H}}\left[ s \right]{{\mathbf{u}}_{ki}}\left[ {m - s} \right]{x_i}} \left[ m \right] + {\mathbf{w}_k}\left[ s \right] ,
\end{align}
where
\begin{align}
{{\mathbf{a}}_k}\left[ s \right] &\triangleq {\left[ {{a_{k1}}\left[ s \right], \ldots, {a_{kL}}\left[ s \right]} \right]^{\text{T}}} \in {\mathbb{C}^L} , \notag \\
{{\mathbf{w}}_k}\left[ s \right] &\triangleq \sum\limits_{l = 1}^L {a_{kl}^*\left[ s \right]} {\mathbf{v}}_{kl}^{\text{H}}\left[ s \right]{{\mathbf{w}}_l}\left[ s \right], \notag \\
{{\mathbf{u}}_{ki}}\left[ {m - s} \right] &\triangleq \left[ {{\mathbf{v}}_{k1}^{\text{H}}\left[ s \right]\left( {{I_{i1}}\left[ {m - s} \right]{{{\mathbf{\bar h}}}_{i1}} + {I_{D}}\left[ {m - s} \right]{{\mathbf{h}}_{i1}}} \right)} \right. , \notag \\
  &\;\;\;\;\;\;\;\;\left. { \ldots, {\mathbf{v}}_{kL}^{\text{H}}\left[ s \right]\left( {{I_{iL}}\left[ {m - s} \right]{{{\mathbf{\bar h}}}_{iL}} + {I_{D}}\left[ {m - s} \right]{{\mathbf{h}}_{iL}}} \right)} \right] \in {\mathbb{C}^L} . \notag
\end{align}
\begin{lemm}
An achievable SE of TA $k$ is ${\mathrm{S}}{{\mathrm{E}}_k^\mathrm{lo}}\left[ s \right] = {\log _2}\left( {1 + \mathrm{SINR}_k^\mathrm{lo}\left[s\right]} \right)$ with $\mathrm{SINR}_k^\mathrm{lo}\left[s\right]$ is given as
\begin{align}\label{SINR_lo}
\frac{{{p_k}{{\left| {{\mathbf{a}}_k^{\mathrm{H}}\left[ s \right]\mathbb{E}\left\{ {{{\mathbf{u}}_{kk}}\left[ 0 \right]} \right\}} \right|}^2}}}{{\sum\limits_{i = 1}^K {{p_i}\sum\limits_{m = 1}^M {\mathbb{E}\left\{ {{{\left| {{\mathbf{a}}_k^{\mathrm{H}}\left[ s \right]{{\mathbf{u}}_{ki}}\left[ {m - s} \right]} \right|}^2}} \right\} - {p_k}{{\left| {{\mathbf{a}}_k^{\mathrm{H}}\left[ s \right]\mathbb{E}\left\{ {{{\mathbf{u}}_{kk}}\left[ 0 \right]} \right\}} \right|}^2} + {\sigma ^2}{\mathbf{a}}_k^{\mathrm{H}}\left[ s \right]{{\mathbf{\ddot \Lambda }}_k}\left[ s \right]{{\mathbf{a}}_k}\left[ s \right]} } }} ,
\end{align}
where
\begin{align}
{{\mathbf{\ddot \Lambda }}_k}\left[ s \right] \triangleq {\mathrm{diag}}\left( {\mathbb{E}\left\{ {{{\left\| {{{\mathbf{v}}_{k1}}\left[ s \right]} \right\|}^2}} \right\}, \ldots ,\mathbb{E}\left\{ {{{\left\| {{{\mathbf{v}}_{kL}}\left[ s \right]} \right\|}^2}} \right\}} \right) \in {\mathbb{C}^{L \times L}}.
\end{align}
\end{lemm}
\begin{IEEEproof}
It follows from \cite[Proposition 2]{bjornson2019making} holding for any combining scheme.
\end{IEEEproof}
\begin{cor}
The effective SINR for TA $k$ is maximized by
\begin{align}
{{\mathbf{a}}_k}\left[ s \right] = {\left( {\sum\limits_{i = 1}^K {{p_i}\sum\limits_{m = 1}^M {\mathbb{E}\left\{ {{{\mathbf{u}}_{ki}}\left[ {m - s} \right]{\mathbf{u}}_{ki}^{\mathrm{H}}\left[ {m - s} \right]} \right\} + {\sigma ^2}{{\mathbf{\ddot \Lambda }}_k}\left[ s \right]} } } \right)^{ - 1}}\mathbb{E}\left\{ {{{\mathbf{u}}_{kk}}\left[ 0 \right]} \right\},
\end{align}
which leads to the maximum value
\begin{align}
  {\mathrm{S}}{{\mathrm{E}}_k^\mathrm{lsfd}}\left[ s \right] &= {\log _2}\left( {1 + {p_k}\mathbb{E}\left\{ {{\mathbf{u}}_{kk}^{\mathrm{H}}\left[ 0 \right]} \right\}\left( {\sum\limits_{i = 1}^K {{p_i}\sum\limits_{m = 1}^M {\mathbb{E}\left\{ {{{\mathbf{u}}_{ki}}\left[ {m - s} \right]{\mathbf{u}}_{ki}^{\mathrm{H}}\left[ {m - s} \right]} \right\}} } } \right.} \right. \notag \\
  &\;\;\;\;\;\;\;\;\;\;\;\;\;\;\;\;\left. {{{\left. { + {\sigma ^2}{{\mathbf{\ddot \Lambda }}_k}\left[ s \right] - {p_k}\mathbb{E}\left\{ {{{\mathbf{u}}_{kk}}\left[ 0 \right]} \right\}\mathbb{E}\left\{ {{\mathbf{u}}_{kk}^{\mathrm{H}}\left[ 0 \right]} \right\}} \right)}^{ - 1}}\mathbb{E}\left\{ {{{\mathbf{u}}_{kk}}\left[ 0 \right]} \right\}} \right) .
\end{align}
\end{cor}
\begin{IEEEproof}
A process similar to Corollary \ref{cor1}, the structure of \eqref{SINR_lo} allows computing the deterministic weighting vector ${{\mathbf{a}}_k}\left[s\right]$ that maximizes SINR.
\end{IEEEproof}
In the following, we derive closed-form SE expressions with the MR combining ${{{\mathbf{v}}_{kl}}\left[ s \right]} = {{{\mathbf{\hat g}}_{kl}}}$.
In order to detect the symbol from the $k$th TA at $s$th subcarrier, the $l$th AP multiplies the conjugate of channel estimate by the received signal. The derived quantity is sent to the CPU through the fronthaul with the weight coefficients. We then can divide the received signal into five parts as
\begin{align}\label{y_k}
  {{\hat y}_k}\left[ s \right] &= \sum\limits_{l = 1}^L {a_{kl}^*\left[ s \right]{{\overset{\lower0.5em\hbox{$\smash{\scriptscriptstyle\smile}$}}{y} }_{kl}}\left[ s \right]} =  \sum\limits_{l = 1}^L {a_{kl}^*\left[ s \right]{\mathbf{\hat g}}_{kl}^{\text{H}}{{\mathbf{y}}_l}\left[ s \right]} \notag \\
   &= \underbrace {\sum\limits_{l = 1}^L {\sqrt {{p_k}} a_{kl}^*\left[ s \right]\mathbb{E}\left\{ {{\mathbf{\hat g}}_{kl}^{\text{H}}\left( {{I_{kl}}\left[ 0 \right]{{{\mathbf{\bar h}}}_{kl}} + {I_{D}}\left[ 0 \right]{{\mathbf{h}}_{kl}}} \right)} \right\}} {x_k}\left[ s \right]}_{{\text{D}}{{\text{S}}_k}\left[ s \right]} + \underbrace {\sum\limits_{l = 1}^L {a_{kl}^*\left[ s \right]{\mathbf{\hat g}}_{kl}^{\text{H}}{{\mathbf{w}}_l}\left[ s \right]} }_{{\text{N}}{{\text{S}}_k}\left[ s \right]} \notag \\
   &+ \underbrace {\sum\limits_{l = 1}^L {\sqrt {{p_k}} a_{kl}^*\left[ s \right]} \left( {{\mathbf{\hat g}}_{kl}^{\text{H}}\left( {{I_{kl}}\left[ 0 \right]{{{\mathbf{\bar h}}}_{kl}} + {I_{D}}\left[ 0 \right]{{\mathbf{h}}_{kl}}} \right) - \mathbb{E}\left\{ {{\mathbf{\hat g}}_{kl}^{\text{H}}\left( {{I_{kl}}\left[ 0 \right]{{{\mathbf{\bar h}}}_{kl}} + {I_{D}}\left[ 0 \right]{{\mathbf{h}}_{kl}}} \right)} \right\}} \right){x_k}\left[ s \right]}_{{\text{B}}{{\text{U}}_k}\left[ s \right]} \notag \\
   &+ \sum\limits_{m \ne s}^M {\underbrace {\sum\limits_{l = 1}^L {\sqrt {{p_k}} a_{kl}^*\left[ s \right]{\mathbf{\hat g}}_{kl}^{\text{H}}\left( {{I_{kl}}\left[ {m - s} \right]{{{\mathbf{\bar h}}}_{kl}} + {I_{D}}\left[ {m - s} \right]{{\mathbf{h}}_{kl}}} \right){x_k}\left[ m \right]} }_{{\text{IC}}{{\text{I}}_k}\left[ {m - s} \right]}}  \notag \\
   &+ \sum\limits_{i \ne k}^K {\sum\limits_{m = 1}^M {\underbrace {\sum\limits_{l = 1}^L {\sqrt {{p_i}} a_{kl}^*\left[ s \right]{\mathbf{\hat g}}_{kl}^{\text{H}}\left( {{I_{il}}\left[ {m - s} \right]{{{\mathbf{\bar h}}}_{il}} + {I_{D}}\left[ {m - s} \right]{{\mathbf{h}}_{il}}} \right){x_i}\left[ m \right]} }_{{\text{U}}{{\text{I}}_{ki}}\left[ {m - s} \right]}} }  ,
\end{align}
where ${{\text{D}}{{\text{S}}_k}\left[ s \right]}$ represents the desired signal at $s$th subcarrier, ${{\text{B}}{{\text{U}}_k}\left[ s \right]}$ is the beamforming gain uncertainty at $s$th subcarrier, ${{\text{IC}}{{\text{I}}_k}\left[ {m - s} \right]}$ represents the ICI from $m$th subcarrier, ${{\text{U}}{{\text{I}}_{ki}}\left[ {m - s} \right]}$ denotes the interference resulted from transmitted data from other TAs, and ${{\text{N}}{{\text{S}}_k}\left[ s \right]}$ represents the noise term, respectively.
\begin{thm}\label{thm3}
Based on the signal in \eqref{y_k}, the capacity of TA $k$ at $s$th subcarrier is
\begin{align}\label{LSFD_MR}
{\mathrm{S}}{{\mathrm{E}}_k^\mathrm{lo,mr}}\left[ s \right] = {\log _2}\left( {1 + \mathrm{SINR}_k^\mathrm{lo,mr}\left[s\right]} \right)
\end{align}
with $\mathrm{SINR}_k^\mathrm{lo,mr}\left[s\right]$ is given as
\begin{align}\label{SINR_lomr}
  \mathrm{SINR}_k^\mathrm{lo,mr}\left[s\right] &= {p_k}{\left| {{\mathbf{a}}_k^{\mathrm{H}}\left[ s \right]{{\mathbf{b}}_k}} \right|^2}\left( {\sum\limits_{i = 1}^K {{p_i}\sum\limits_{m = 1}^M  {\mathbf{a}}_k^{\mathrm{H}}\left[ s \right]{{\mathbf{\Xi }}_{ki}}\left[ {m - s} \right]{{\mathbf{a}}_k}\left[ s \right]}  + {p_k}\sum\limits_{m \ne s}^M  {{\left| {{\mathbf{a}}_k^{\mathrm{H}}\left[ s \right]{{\mathbf{c}}_k}\left[ {m - s} \right]} \right|}^2}} \right. \notag \\
  &{\left. { + \sum\limits_{i \ne k}^K {{p_i}\sum\limits_{m = 1}^M  {{\left| {{\mathbf{a}}_k^{\mathrm{H}}\left[ s \right]{{\mathbf{d}}_{ki}}\left[ {m - s} \right]} \right|}^2}}  + {\sigma ^2}{\mathbf{a}}_k^{\mathrm{H}}\left[ s \right]{{\mathbf{\Lambda }}_k}{{\mathbf{a}}_k}\left[ s \right]} \right)^{ - 1}} ,
\end{align}
where
\begin{align}
&{{\mathbf{b}}_k} \triangleq {\left[ {\left( {{I_{k1}}\left[ 0 \right]{\mathbf{\bar h}}_{k1}^{\mathrm{H}}{{{\mathbf{\bar h}}}_{k1}} + {I_{{D}}}\left[ 0 \right]{\mathrm{tr}}\left( {{{\mathbf{Q}}_{k1}}} \right)} \right), \ldots, \left( {{I_{kL}}\left[ 0 \right]{\mathbf{\bar h}}_{kL}^{\mathrm{H}}{{{\mathbf{\bar h}}}_{kL}} + {I_{D}}\left[ 0 \right]{\mathrm{tr}}\left( {{{\mathbf{Q}}_{kL}}} \right)} \right)} \right]^{\mathrm{T}}} \in {\mathbb{C}^L} , \notag \\
&{{\mathbf{\Xi }}_{ki}}\!\left[ {m \!-\! s} \right] \triangleq {\mathrm{diag}}\left( {\left( {I_{D}^2\left[ {m \!-\! s} \right]{\mathrm{tr}}\left( {{{\mathbf{R}}_{i1}}{{\mathbf{Q}}_{k1}}} \right) \!+\! I_{D}^2\left[ {m \!-\! s} \right]{\mathbf{\bar h}}_{k1}^{\mathrm{H}}{{\mathbf{R}}_{i1}}{{{\mathbf{\bar h}}}_{k1}} \!+\! I_{i1}^2\left[ {m \!-\! s} \right]{\mathbf{\bar h}}_{i1}^{\mathrm{H}}{{\mathbf{Q}}_{k1}}{{{\mathbf{\bar h}}}_{i1}}} \right),} \right. \notag \\
  &\;\;\;\;\;\left. { \ldots ,\left( {I_{D}^2\!\left[ {m \!-\! s} \right]{\mathrm{tr}}\left( {{{\mathbf{R}}_{iL}}{{\mathbf{Q}}_{kL}}} \right) \!+\! I_{D}^2\left[ {m \!-\! s} \right]{\mathbf{\bar h}}_{kL}^{\mathrm{H}}{{\mathbf{R}}_{iL}}{{{\mathbf{\bar h}}}_{kL}} \!+\! I_{iL}^2\!\left[ {m \!-\! s} \right]{\mathbf{\bar h}}_{iL}^{\mathrm{H}}{{\mathbf{Q}}_{kL}}{{{\mathbf{\bar h}}}_{iL}}} \right)} \right) \in {\mathbb{C}^{L \times L}} , \notag\\
  &{{\mathbf{c}}_k}\left[ {m - s} \right] \triangleq \left[ {\left( {{I_{k1}}\left[ {m - s} \right]{\mathbf{\bar h}}_{k1}^{\mathrm{H}}{{{\mathbf{\bar h}}}_{k1}} + {I_{D}}\left[ {m - s} \right]{\mathrm{tr}}\left( {{{\mathbf{Q}}_{k1}}} \right)} \right)} \right. , \notag \\
  &\;\;\;\;\;\;\;\;\;\;\;\;\;\;\;\;\;\;\;\;\;\;\;{\left. { \ldots, \left( {{I_{kL}}\left[ {m - s} \right]{\mathbf{\bar h}}_{kL}^{\mathrm{H}}{{{\mathbf{\bar h}}}_{kL}} + {I_{D}}\left[ {m - s} \right]{\mathrm{tr}}\left( {{{\mathbf{Q}}_{kL}}} \right)} \right)} \right]^{\mathrm{T}}} \in {\mathbb{C}^L} , \notag \\
  &{{\mathbf{d}}_{ki}}\left[ {m - s} \right] \triangleq \left[ {{I_{i1}}\left[ {m - s} \right]{\mathbf{\bar h}}_{k1}^{\mathrm{H}}{{{\mathbf{\bar h}}}_{i1}}} \right.{\left. { ,\ldots, {I_{iL}}\left[ {m - s} \right]{\mathbf{\bar h}}_{kL}^{\mathrm{H}}{{{\mathbf{\bar h}}}_{iL}}} \right]^{\mathrm{T}}} \in {\mathbb{C}^L} , \notag \\
  &{{\mathbf{ \Lambda }}_k} \triangleq {\mathrm{diag}}\left( {\left( {{\mathrm{tr}}\left( {{{\mathbf{Q}}_{k1}}} \right) + {\mathbf{\bar h}}_{k1}^{\mathrm{H}}{{{\mathbf{\bar h}}}_{k1}}} \right), \ldots ,\left( {{\mathrm{tr}}\left( {{{\mathbf{Q}}_{kL}}} \right) + {\mathbf{\bar h}}_{kL}^{\mathrm{H}}{{{\mathbf{\bar h}}}_{kL}}} \right)} \right) \in {\mathbb{C}^{L \times L}} . \notag
\end{align}
\end{thm}
\begin{IEEEproof}
Please refer to Appendix B.
\end{IEEEproof}
The increase of the DFO effect leads to greater ICI interference coefficients ($m \!\ne\! s$), which makes SINR smaller. The SE in Theorem \ref{thm3} is a general expression of CF massive MIMO including ICI effect, LoS link, spatial correlation and multiple antennas. By making ${\varepsilon _{il}} \!=\! 0$, $\bar K_{kl} \!=\! 0$ and ${{\mathbf{R}}_{kl}} \!=\! \beta _{kl}^{{\text{NLoS}}}{{\mathbf{I}}_N}$, we can obtain various special expressions in previous CF studies.
\begin{cor}\label{LSFD1}
Using the LSFD weights, the effective SINR of TA $k$ at $s$th subcarrier is maximized by
\begin{align}\label{LSFD2}
  {{\mathbf{a}}_k}\left[ s \right] &= \left( {\sum\limits_{i = 1}^K {{p_i}\sum\limits_{m = 1}^M {{{\mathbf{\Xi }}_{ki}}\left[ {m - s} \right]} }  + {p_k}\sum\limits_{m \ne s}^M {{{\mathbf{c}}_k}\left[ {m - s} \right]{\mathbf{c}}_k^{\mathrm{H}}\left[ {m - s} \right]} } \right. \notag \\
   &\;\;\;\;\;\;\;\;\;\;\;\;\;\;\;\;\;\;\;\;\;\;\;+ {\sum\limits_{i \ne k}^K {\left. {{p_i}\sum\limits_{m \ne s}^M {{{\mathbf{d}}_{ki}}\left[ {m - s} \right]{\mathbf{d}}_{ki}^{\mathrm{H}}\left[ {m - s} \right]}  + {\sigma ^2}{{\mathbf{\Lambda }}_k}} \right)} ^{ - 1}}{{\mathbf{b}}_k} ,
\end{align}
which leads to the maximum SE as
\begin{align}
  {\mathrm{SE}}_k^{{\mathrm{lsfd,mr}}}\left[ s \right] &= {\log _2}\left( {1 + {p_k}{\mathbf{b}}_k^{\mathrm{H}}\left( {\sum\limits_{i = 1}^K {{p_i}\sum\limits_{m = 1}^M {{{\mathbf{\Xi }}_{ki}}\left[ {m - s} \right]} } } \right.} \right. \notag \\
  &\!\!\!\!\!\!\left. {{{\left. { + {p_k}\sum\limits_{m \ne s}^M {{{\mathbf{c}}_k}\left[ {m \!-\! s} \right]{\mathbf{c}}_k^{\mathrm{H}}\left[ {m \!-\! s} \right]} \!+\! \sum\limits_{i \ne k}^K {{p_i}\sum\limits_{m \ne s}^M {{{\mathbf{d}}_{ki}}\left[ {m \!-\! s} \right]{\mathbf{d}}_{ki}^{\mathrm{H}}\left[ {m \!-\! s} \right]} \!+\! {\sigma ^2}{{\mathbf{\Lambda }}_k}} } \right)}^{ - 1}}{{\mathbf{b}}_k}} \right) .
\end{align}
\end{cor}
\begin{IEEEproof}
It follows similar steps in Corollary \ref{cor1}.
\end{IEEEproof}
If we want to reduce the complexity of LSFD, then the conventional MF receiver cooperation is obtained by using equal weights $\mathbf{a}_{k}[s]=[1 / L, \ldots, 1 / L]^{\mathrm{T}}$.

\subsection{Small Cell Systems With OFDM}

In this section, we compare with a conventional small cell system that consists of $L$ APs and $K$ TAs, which has the same locations as in the CF case but each TA is only served by one AP, which gives the highest SE performance. An arbitrary receive combining vector based on the channel estimate is used to multiply the received signal for detecting the desired signal.
\begin{thm}
Then the capacity of TA $k$ is lower bounded by
\begin{align}
{\mathrm{S}}{{\mathrm{E}}_k^\mathrm{sc}}\left[ s \right] = \mathop {\max }\limits_{l \in \left\{ {1, \ldots ,L} \right\}} \mathbb{E}\left\{ {{{\log }_2}\left( {1 + \mathrm{SINR}_{kl}^\mathrm{sc}\left[s\right] } \right)} \right\} ,
\end{align}
where $\mathrm{SINR}_{kl}^\mathrm{sc}\left[s\right]$ is given as
\begin{align}\label{smallcell}
 \mathrm{SINR}_{kl}^\mathrm{sc}\left[s\right] &= {p_k}{\left| {{\mathbf{v}}_{kl}^{\mathrm{H}}\left[ s \right]{{\mathbf{f}}_{kl}}\left[ 0 \right]} \right|^2}\left( {{p_k}\sum\limits_{m \ne s}^M {{{\left| {{\mathbf{v}}_{kl}^{\mathrm{H}}\left[ s \right]{{\mathbf{f}}_{kl}}\left[ {m - s} \right]} \right|}^2}}  + } \right.\sum\limits_{i \ne k}^K {{p_i}\sum\limits_{m = 1}^M {{{\left| {{\mathbf{v}}_{kl}^{\mathrm{H}}\left[ s \right]{{\mathbf{f}}_{il}}\left[ {m - s} \right]} \right|}^2}} }  \notag \\
  &{\left. { + {\mathbf{v}}_{kl}^{\mathrm{H}}\left[ s \right]\left( {\sum\limits_{i = 1}^K {{p_i}\sum\limits_{m = 1}^M {I_D^2\left[ {m - s} \right]{{\mathbf{C}}_{il}} + {\sigma ^2}{{\mathbf{I}}_N}} } } \right){{\mathbf{v}}_{kl}}\left[ s \right]} \right)^{ - 1}} .
\end{align}
\end{thm}
\begin{IEEEproof}
It follows similar steps in Lemma \ref{thm1}.
\end{IEEEproof}
It is worthy noting that the maximum value in \eqref{smallcell} is achieved with the MMSE combining in \eqref{v_kl} and is given by
\begin{align}
  {\text{SINR}}_{kl}^{{\text{sc,mmse}}}\left[ s \right] &= {p_k}{\mathbf{f}}_{kl}^{\text{H}}\left[ 0 \right]\left( {{p_k}\sum\limits_{m \ne s}^M {{{\mathbf{f}}_{kl}}\left[ {m - s} \right]{\mathbf{f}}_{kl}^{\text{H}}\left[ {m - s} \right]} } \right. + \sum\limits_{i \ne k}^K {{p_i}\sum\limits_{m = 1}^M {{{\mathbf{f}}_{il}}\left[ {m - s} \right]{\mathbf{f}}_{il}^{\text{H}}\left[ {m - s} \right]} }  \notag \\
  &{\left. { + \sum\limits_{i = 1}^K {{p_i}\sum\limits_{m = 1}^M {I_D^2\left[ {m - s} \right]{{\mathbf{C}}_{il}} + {\sigma ^2}{{\mathbf{I}}_N}} } } \right)^{ - 1}}{{\mathbf{f}}_{kl}}\left[ 0 \right] .
\end{align}
The MR combining with $\mathbf{v}_{kl} = \mathbf{\hat g}_{kl}$ also can be used to obtain a suboptimal value of \eqref{smallcell}.
\subsection{Cellular Systems With OFDM}

We consider a cellular network with one cell and $L_\text{bs} = LN$ antennas at the cellular BS, which is like a special case of CF with $L=1$ and $N$ is the total number of antennas.
The block-fading channel from BS to TA $k$ is modeled as ${\mathbf{g}}_k^{\text{c}} \sim \mathcal{C}\mathcal{N}\left( {{\mathbf{\bar h}}_k^{\text{c}},{\mathbf{R}}_k^{\text{c}}} \right)$, where ${\mathbf{\bar h}}_k^{\text{c}} \in {\mathbb{C}^{LN}}$ is the LoS component, and ${\mathbf{R}}_k^{\text{c}} \in {\mathbb{C}^{LN \times LN}}$ is the spatial correlation matrix of the NLoS components.
The MMSE estimate ${\mathbf{\hat g}}_k^{\text{c}} \in {\mathbb{C}^{LN}}$ and the independent estimation error ${\mathbf{\tilde g}}_k^{\text{c}} \in {\mathbb{C}^{LN}}$ are respectively given by ${\mathbf{\hat g}}_k^{\text{c}} \sim \mathcal{C}\mathcal{N}\left( {{\mathbf{\bar h}}_k^{\text{c}},{\mathbf{Q}}_k^{\text{c}}} \right)$ and ${\mathbf{\tilde g}}_k^{\text{c}} = {\mathbf{g}}_k^{\text{c}} - {\mathbf{\hat g}}_k^{\text{c}} \sim \mathcal{C}\mathcal{N}\left( {{\mathbf{0}},{\mathbf{C}}_k^{\text{c}}} \right)$, where $ {\mathbf{Q}}_k^{\text{c}} = {p_k}{\tau _p}{\mathbf{R}}_k^{\text{c}}{\mathbf{\Psi }}_k^{\text{c}}{\mathbf{R}}_k^{\text{c}} $ and ${\mathbf{\Psi }}_k^{\text{c}} = {\left( {{p_k}{\tau _p}{\mathbf{R}}_k^{\text{c}} + {\sigma ^2}{{\mathbf{I}}_N}} \right)^{ - 1}} $.
An arbitrary receive combining vector based on the channel estimate is used to multiply the received signal for detecting the desired signal.
\begin{thm}
Then, the SE of TA $k$ is given by
\begin{align}
{\mathrm{S}}{{\mathrm{E}}_k^\mathrm{c}}\left[ s \right] = \mathbb{E}\left\{ {{{\log }_2}\left( {1 + \mathrm{SINR}_k^\mathrm{c}\left[s\right]} \right)} \right\} ,
\end{align}
with $\mathrm{SINR}_k^\mathrm{c}\left[s\right]$ is given as
\begin{align}\label{cellular}
  \mathrm{SINR}_k^\mathrm{c}\left[s\right] &\!=\! {p_k}{\left| {{{\left( {{\mathbf{v}}_k^{\mathrm{c}}\left[ s \right]} \right)}^{\mathrm{H}}}{\mathbf{f}}_k^{\mathrm{c}}\left[ 0 \right]} \right|^2}\left( {{p_k}\sum\limits_{m \ne s}^M {{{\left| {{{\left( {{\mathbf{v}}_k^{\mathrm{c}}\left[ s \right]} \right)}^{\mathrm{H}}}{\mathbf{f}}_k^{\mathrm{c}}\left[ {m \!-\! s} \right]} \right|}^2} \!+\! \sum\limits_{i \ne k}^K {{p_i}\sum\limits_{m = 1}^M {{{\left| {{{\left( {{\mathbf{v}}_k^{\mathrm{c}}\left[ s \right]} \right)}^{\mathrm{H}}}{\mathbf{f}}_i^{\mathrm{c}}\left[ {m \!-\! s} \right]} \right|}^2}} } } } \right. \notag \\
  &{\left. { + {{\left( {{\mathbf{v}}_k^{\mathrm{c}}\left[ s \right]} \right)}^{\mathrm{H}}}\left( {\sum\limits_{i = 1}^K {{p_i}\sum\limits_{m = 1}^M {I_D^2\left[ {m - s} \right]{\mathbf{C}}_i^{\mathrm{c}} + {\sigma ^2}{{\mathbf{I}}_{LN}}} } } \right){\mathbf{v}}_k^{\mathrm{c}}\left[ s \right]} \right)^{ - 1}} ,
\end{align}
where
\begin{align}
\mathbf{f}_i^\mathrm{c}\left[m-s\right] =  {{I_i}\left[ {m - s} \right]{\mathbf{\bar h}}_i^{\mathrm{c}} + {I_{D}}\left[ {m - s} \right]{\mathbf{\hat h}}_i^{\mathrm{c}}} .
\end{align}
\end{thm}
\begin{IEEEproof}
It follows similar steps in Lemma \ref{thm1}.
\end{IEEEproof}
Moreover, utilizing the MMSE combining
\begin{align}
{\mathbf{v}}_k^{\text{c}}\left[s\right] = {p_k}{\left( {\sum\limits_{i = 1}^K {{p_i}\sum\limits_{m = 1}^M {\left( {{\mathbf{f}}_i^{\text{c}}\left[ {m - s} \right]{{\left( {{\mathbf{f}}_i^{\text{c}}\left[ {m - s} \right]} \right)}^{\text{H}}} + I_D^2\left[ {m - s} \right]{\mathbf{C}}_i^{\text{c}}} \right) + {\sigma ^2}{{\mathbf{I}}_{LN}}} } } \right)^{ - 1}}{\mathbf{f}}_k^{\text{c}}\left[ 0 \right] ,
\end{align}
we can derive the maximum value of \eqref{cellular} as
\begin{align}
  {\text{SINR}}_k^{{\text{c}},{\text{mmse}}}\left[ s \right] &\!=\! {p_k}{\left( {{\mathbf{f}}_k^{\text{c}}\left[ 0 \right]} \right)^{\text{H}}}\left( {{p_k}\sum\limits_{m \ne s}^M {{\mathbf{f}}_k^{\text{c}}\left[ {m \!-\! s} \right]{{\left( {{\mathbf{f}}_k^{\text{c}}\left[ {m \!-\! s} \right]} \right)}^{\text{H}}}}  \!+\! } \right.\sum\limits_{i \ne k}^K {{p_i}\sum\limits_{m = 1}^M {{\mathbf{f}}_i^{\text{c}}\left[ {m \!-\! s} \right]{{\left( {{\mathbf{f}}_i^{\text{c}}\left[ {m \!-\! s} \right]} \right)}^{\text{H}}}} }  \notag \\
  &{\left. { + \sum\limits_{i = 1}^K {{p_i}\sum\limits_{m = 1}^M {I_D^2\left[ {m - s} \right]{\mathbf{C}}_i^{\text{c}} + {\sigma ^2}{{\mathbf{I}}_{LN}}} } } \right)^{ - 1}}{\mathbf{f}}_k^{\text{c}}\left[ 0 \right] .
\end{align}
In addition, the MR combining ${\mathbf{v}}_k^{\text{c}}\left[s\right] = {\mathbf{\hat g}}_k^{\text{c}}$ can be used to obtain a suboptimal value of \eqref{cellular}.

\section{Practical Application in HST Communications}\label{se:Scalable}

According to the analysis in Section \ref{se:performance}, we find that the two main problems faced in the HST communication are the scalable deployment of wireless networks and the performance loss caused by DFO.
In this section, we investigate the TA-centric CF massive MIMO-OFDM system in HST communications, where each TA is served by the AP subset providing the best channel conditions.
Besides, we focus on power control to reduce the impact of DFO.

\subsection{TA-centric CF Massive MIMO-OFDM}

In our considered HST communications, there are thousands of APs along the side of the entire track, but the train does not need to be served by all of them. As in the traditional CF architecture, it is impossible for TAs to be served by all APs, due to the huge computational complexity and fronthaul requirements. TA-centric dynamic cooperation clustering is key to realize the practical application of CF massive MIMO-OFDM in the HST scenario.
The idea is to have each TA select the APs that give the best performance, that is each TA is jointly served by only a subset of nearby APs. This is reasonable because APs far away from TAs have little or negligible impact on performance.

For achieving dynamic cooperation clustering in the system, we define a set of diagonal matrices ${{\mathbf{D}}_{il}} \in {\mathbb{C}^{N \times N}}$ for $i = 1, \ldots , K$ and $l = 1, \ldots ,L$ to determine which AP antennas may transmit to which TAs. Moreover, we define ${\mathcal{M}_i} \in \left\{ {1, \ldots ,L} \right\}$ as the subset of APs served by TA $i$ and ${\mathcal{D}_l} \in \left\{ {1, \ldots ,K} \right\}$ as the subset of TAs served by AP $l$. Then, we provide the dynamic cluster formation scheme in Algorithm \ref{cluster_formation} to obtain the TA-centric settings as
\begin{align}\label{D_il}
{{\mathbf{D}}_{il}} = \left\{ {\begin{array}{*{20}{c}}
  {{{\mathbf{I}}_N},l \in {\mathcal{M}_i}} ,\\
  {{{\mathbf{0}}_N},l \notin {\mathcal{M}_i}} .
\end{array}} \right.
\end{align}
\begin{algorithm}[tb]
\caption{Dynamic cluster formation scheme.}
\label{cluster_formation}
\begin{algorithmic}[1]
\Require
the time interval $n$, the large-scale fading $\zeta_{i l}$, the threshold $\Theta $ [dB];
\Ensure
the TA-centric settings ${{\mathbf{D}}_{il}}$;
\For{each $n$}
\State the accessing TA $i$ measures the large-scale fading $\zeta_{i l}$;
\State the TA $i$ appoints AP $\ell$ with $\ell=\arg \max _{l} \zeta_{il}$ as its master AP;
\State compute ${{\bar \Theta }_{il}} = {\zeta _{i\ell }} - {\zeta _{il}}$ [dB];
\If{${{\bar \Theta }_{il}} < \Theta$}
\State  ${{\mathbf{D}}_{il}} = {{\mathbf{I}}_N}$;
\EndIf
\If{${{\bar \Theta }_{il}} \geqslant \Theta$}
\State  ${{\mathbf{D}}_{il}} = {{\mathbf{0}}_N}$;
\EndIf
\State \Return {the TA-centric settings ${{\mathbf{D}}_{il}}$.}
\EndFor
\end{algorithmic}
\end{algorithm}
Note that the original CF massive MIMO is the special case of ${{\mathbf{D}}_{il}} = {{\mathbf{I}}_N}$ for $\forall i$ and $\forall l$, where all AP antennas serve all TAs.
For CF massive MIMO with local processing, AP $l$ can locally select the combiner $\mathbf{v}_{kl}$ on the basis of its local channel estimates $\left\{ {{{{\mathbf{\hat g}}}_{il}}:i \in {\mathcal{D}_l}} \right\}$.
The collective channel estimate ${{{{\mathbf{\hat g}}}_{il}}}$ can be only partially computed since only some channel estimate of TAs is needed for AP $l$ to combining. More precisely, the AP $l$ knows the collective channel estimate and the collective estimation error as
\begin{align}
{{\mathbf{D}}_{kl}}{{{\mathbf{\hat g}}}_{kl}} &\sim \mathcal{C}\mathcal{N}\left( {{{\mathbf{D}}_{kl}}{{{\mathbf{\bar h}}}_{kl}},{{\mathbf{D}}_{kl}}{{\mathbf{Q}}_{kl}}{{\mathbf{D}}_{kl}}} \right) , \\
{{\mathbf{D}}_{kl}}{{{\mathbf{\tilde g}}}_{kl}} &= {{\mathbf{D}}_{kl}}{{\mathbf{g}}_{kl}} - {{\mathbf{D}}_{kl}}{{{\mathbf{\hat g}}}_{kl}} \sim \mathcal{C}\mathcal{N}\left( {{\mathbf{0}},{{\mathbf{D}}_{kl}}{{\mathbf{C}}_{kl}}{{\mathbf{D}}_{kl}}} \right).
\end{align}
Using MR combining $\mathbf{v}_{kl} = {{{\mathbf{D}}_{kl}}{{{\mathbf{\hat g}}}_{kl}}}$, the AP then computes its local estimate of $x_k$ as ${{\overset{\lower0.5em\hbox{$\smash{\scriptscriptstyle\smile}$}}{y} }_{kl}}\left[ s \right] = {\left( {{{\mathbf{D}}_{kl}}{{{\mathbf{\hat g}}}_{kl}}} \right)^{\text{H}}}{{\mathbf{y}}_l}\left[ s \right]$.
The local estimates of all APs that serve TA $k$ are then sent to a CPU where the final estimate of $x_k$ is obtained by taking the sum of the local estimates:
\begin{align}\label{CPU}
{{\hat y}_k}\left[ s \right] = \sum\limits_{l = 1}^L {a_{kl}^*\left[ s \right]{{\overset{\lower0.5em\hbox{$\smash{\scriptscriptstyle\smile}$}}{y} }_{kl}}\left[ s \right]} = \sum\limits_{l = 1}^L {a_{kl}^*\left[ s \right]{\left( {{{\mathbf{D}}_{kl}}{{{\mathbf{\hat g}}}_{kl}}} \right)^{\text{H}}}{{\mathbf{y}}_l}\left[ s \right]}.
\end{align}
Based on the received signal \eqref{CPU} at the TA-centric settings, an achievable uplink SE for TA $k$ can be derived as $\mathrm{SE}_k^\mathrm{tac}\left[s\right]$ following similar steps in Theorem \ref{thm3}, with
\begin{align}\label{generic}
{\mathrm{SINR}}_k^{{\mathrm{tac}}}\left[ s \right] = \frac{{{b_k}\left[ s \right]{p_k}}}{{{\mathbf{f}}_k^{\mathrm{T}}\left[ s \right]{\mathbf{p}} + \sigma _k^2\left[ s \right]}},
\end{align}
where ${\mathbf{p}} = {\left[ {{p_1}, \ldots ,{p_K}} \right]^{\mathrm{T}}}$, ${b_k}\left[ s \right] = {\left| {{\mathbf{a}}_k^{\mathrm{H}}\left[ s \right]{{\mathbf{b}}_k}} \right|^2}$, ${{\mathbf{f}}_k}\left[ s \right] = {\left[ {{f_{k1}}\left[ s \right], \ldots ,{f_{kK}}\left[ s \right]} \right]^{\mathrm{T}}}$, and $\sigma _k^2\left[ s \right] = {\sigma ^2}{\mathbf{a}}_k^{\mathrm{H}}\left[ s \right]{{\mathbf{\Lambda }}_k}{{\mathbf{a}}_k}\left[ s \right]$. Besides, adding the TA-centric settings ${{\mathbf{D}}_{kl}}$ into \eqref{SINR_lomr}, we have
\begin{align}
  {f_{kk}}\left[ s \right] &= \sum\limits_{m = 1}^M {{\mathbf{a}}_k^{\mathrm{H}}\left[ s \right]{{\mathbf{\Xi }}_{kk}}\left[ {m - s} \right]{{\mathbf{a}}_k}\left[ s \right]}  + \sum\limits_{m \ne s}^M {{{\left| {{\mathbf{a}}_k^{\mathrm{H}}\left[ s \right]{{\mathbf{c}}_k}\left[ {m - s} \right]} \right|}^2}},   \\
  {f_{ki}}\left[ s \right] &= \sum\limits_{m = 1}^M {{\mathbf{a}}_k^{\mathrm{H}}\left[ s \right]{{\mathbf{\Xi }}_{ki}}\left[ {m - s} \right]{{\mathbf{a}}_k}\left[ s \right]}  + \sum\limits_{i \ne k}^K {{p_i}\sum\limits_{m = 1}^M {{{\left| {{\mathbf{a}}_k^{\mathrm{H}}\left[ s \right]{{\mathbf{d}}_{ki}}\left[ {m - s} \right]} \right|}^2}} }, i \ne k .
\end{align}
Meanwhile, it is similar as Corollary \ref{LSFD1}, the maximum value of ${\mathrm{S}}{{\mathrm{E}}_k^\mathrm{tac}}\!\left[ s \right]$ can be derived by the LSFD cooperation weights.
Besides, the MF cooperation weights $\mathbf{a}_{k}[s]=[1 / L, \ldots, 1 / L]^{\mathrm{T}}$ can be applied to reduce the complexity of LSFD.
Note that, the form of \eqref{generic} is helpful to solve two power optimization problems: max-min SE fairness and max-sum SE maximization \cite{demir2021foundations}.
\renewcommand{\algorithmicrequire}{\textbf{Initialization:}}
\renewcommand{\algorithmicensure}{\textbf{Output:}}
\begin{algorithm}[tb]
\caption{Fixed-point algorithm.}
\label{FP}
\begin{algorithmic}[1]
\Require
Set arbitrary initial power $\mathbf{p} > \mathbf{0}_K$ and the solution accuracy $\varpi > 0$;
\While{$\mathop {\max }\limits_{k \in \left\{ {1, \ldots ,K} \right\}} {\text{SIN}}{{\text{R}}^\text{tac}_k}\left( {\mathbf{p}} \right) - \mathop {\min }\limits_{k \in \left\{ {1, \ldots ,K} \right\}} {\text{SIN}}{{\text{R}}^\text{tac}_k}\left( {\mathbf{p}} \right) > \varpi $}
\State ${p_k} \leftarrow {p_k}/{\text{SINR}}_k^{{\text{tac}}}\left( {\mathbf{p}} \right),k = 1, \ldots ,K$;
\State ${\mathbf{p}} \leftarrow \left( {{p_{\max }}/\mathop {\max }\limits_{k \in \left\{ {1, \ldots ,K} \right\}} {p_k}} \right){\mathbf{p}}$;
\EndWhile
\Ensure
Optimal transmit powers $\mathbf{p}$.
\end{algorithmic}
\end{algorithm}

\begin{algorithm}[tb]
\caption{Block coordinate descent algorithm.}
\label{BCD}
\begin{algorithmic}[1]
\Require
Set an arbitrary feasible power vector $\mathbf{p}$ and the solution accuracy $\varpi > 0$;
\State The MSE is particularized as ${e_k}\left( {{\mathbf{p}},{u_k}} \right) = u_k^2\left( {{b_k}{p_k} + {\mathbf{f}}_k^{\text{T}}{\mathbf{p}} + \sigma _k^2} \right) - 2{u_k}\sqrt {{b_k}{p_k}}  + 1$;
\While{$\sum\nolimits_{k = 1}^K \!{\left( {{d_k}{e_k}\!\left( {{\mathbf{p}},{u_k}} \right) \!-\! \ln \! \left( {{d_k}} \right)} \right)}$ is either improved more than $\varpi$ or not improved at all}
\State ${u_k} \leftarrow \sqrt {{b_k}{p_k}} /\left( {{b_k}{p_k} + {\mathbf{f}}_k^{\text{T}}{\mathbf{p}} + \sigma _k^2} \right),k = 1, \ldots ,K$;
\State ${d_k} \leftarrow 1/\left( {u_k^2\left( {{b_k}{p_k} + {\mathbf{f}}_k^{\text{T}}{\mathbf{p}} + \sigma _k^2} \right) - 2{u_k}\sqrt {{b_k}{p_k}}  + 1} \right),k = 1, \ldots ,K$;
\State ${p_k} \leftarrow \min \left( {p,{b_k}d_k^2u_k^2/{{\left( {{d_k}u_k^2{b_k} + \sum\nolimits_{i = 1}^K {{d_i}u_i^2{f_{ik}}} } \right)}^2}} \right),k = 1, \ldots ,K$;
\EndWhile
\Ensure
Optimal transmit powers $\mathbf{p}$.
\end{algorithmic}
\end{algorithm}

\subsection{Power Control Schemes}

Different TAs have different DFO in the TA-centric CF network because of their different locations. Therefore, the DFO interference can be reduced by unequal power allocation, exploiting the different propagation conditions of the TAs.

\subsubsection{Fractional Power Control}

Fractional power control is a classical heuristic to mitigate near-far effects in the uplink of CF massive MIMO systems. The transmit power of TA $k$ is ${p_k} = {\eta _k}P$, where $P$ is the maximum transmit power of TAs and $\eta_k$ is the power control coefficient of TA $k$. Based on all the large-scale fading coefficients, the power control coefficient of TA $k$ is given by \cite{bjornson2020scalable} as ${\eta _k} = {{\min \left\{ {{\zeta _i}} \right\}}}/{{{\zeta _k}}},i = 1, \ldots ,K $, where ${\zeta _i} = \sum\limits_{l \in {\mathcal{M}_i}} {{\zeta _{il}}}$.

\subsubsection{Max-min SE Power Control}

Based on \eqref{generic}, the max-min SE optimization problem can be particularized as
\begin{align}\label{mix-min}
  &\mathop { {\text{maximize}}}\limits_{{\mathbf{p}} > {{\mathbf{0}}_K}} \;\mathop {\text{min} }\limits_{k \in \left\{ {1, \ldots ,K} \right\}} \frac{{{b_k}\left[ s \right]{p_k}}}{{{\mathbf{f}}_k^{\text{T}}\left[ s \right]{\mathbf{p}} + \sigma _k^2\left[ s \right]}} \\
  &{\text{subject}}\;{\text{to}}\;\;\;{p_k} < p,k = 1, \ldots ,K. \notag
\end{align}
Due to \eqref{mix-min} meets the conditions in \cite[Lemma 7.1]{demir2021foundations}, the solution to \eqref{mix-min} can be computed by using the fixed-point method in Algorithm \ref{FP}.

\subsubsection{Max-sum SE Power Control}

We will now consider the max-sum SE optimization problem, which is formulated as
\begin{align}\label{max-sum}
  &\mathop { {\text{maximize}}}\limits_{{\mathbf{p}} > {{\mathbf{0}}_K}} \;\sum\limits_{k = 1}^K {{{\log }_2}\left( {1 + \frac{{{b_k}\left[ s \right]{p_k}}}{{{\mathbf{f}}_k^{\text{T}}\left[ s \right]{\mathbf{p}} + \sigma _k^2\left[ s \right]}}} \right)}  \\
  &{\text{subject}}\;{\text{to}}\;\;{p_k} < p,k = 1, \ldots ,K. \notag
\end{align}
It is found \eqref{max-sum} is not convex, however, a local optimum can be attained by the block coordinate descent method given in Algorithm \ref{BCD}.

\section{Numerical Results}\label{se:NR}

We present simulation results to show the validity of our theoretical analysis in this section, and also provide practical insights on the ICI reduction.
We utilize a simulation setup where $L$ APs are deployed on one side of the $d_\text{rai}=1000$ m high-speed railway with an equal distance between each AP, and $K$ TAs are also deployed at equal intervals on the $d_\text{hst}=200$ m HST.
\begin{rem}\label{goal}
Due to many challenges such as high-speed and large penetration loss, a reliable wireless communication system for passengers is still being studied. Note that the LTE-R (railway-dedicated communication system without access for train passengers) with cell architecture can achieve 10 Mbps uplink peak SE with $B = 20$ MHz at $f_c =1.8$ GHz \cite{7553613}. Therefore, 800 Mbps is needed for each car to realize that all 80 users in each car can enjoy reliable services similar to LTE-R.
\end{rem}
\subsection{Parameters and Setup}

Based on the simulation setup in \cite{bjornson2017massive} and \cite{4607239}, we consider communication at the carrier frequency $f_c = 1.8$ GHz, the bandwidth $B=20$ MHz, the transmitted power $p=200$ mW, and the total subcarrier number $M_\text{total}=1024$.
In addition, we assume the coherence time $T_c = 375$ us to support mobility of $v=300$ km/h\footnote{We estimate the channel every time the train has travelled $\lambda/5$, which is frequent enough that the effect of channel aging can be ignored within each coherence block. Besides, Kalman filtering and/or machine learning can be used in the channel prediction for the channel aging reduction \cite{9416909}.}, and the coherence bandwidth $B_c = 120$ kHz with 1250 m path differences. Therefore, each coherence block includes $M=8$ subcarriers, the sampling duration $T_s = 67$ us.
Note that, we use ${\text{SE}} \!=\!\! \left(\! {\sum\nolimits_{k = 1}^K \!{\sum\nolimits_{s = 1}^M {{\text{S}}{{\text{E}}_k}\!\left[ s \right]} } } \right)\!/\!\left( {K\!M} \right)$ to evaluate performance, and define average SE as the average SE at all positions HST travels.

Due to the high probability of LoS link in HST communications, the large-scale fading between the AP $l$ and TA $k$ can be modeled by ${\zeta_{kl}} = \zeta_{0}{\left( {{d_{kl}}} \right)^{ - \alpha }}$, where $\alpha = 3$ is the path loss exponent, and $\zeta_0 = 10^{-12}$ denotes the path loss at a reference distance (e.g., $d_{kl} = 1$ km) between the transmitter and the receiver \cite{7335646}.
We assume each AP is equipped with a uniform linear array with half-wavelength antenna spacing. We then model the LoS response and the spatial correlation matrix for Rician fading from TA $k$ to AP $l$ as  \cite{bjornson2017massive}:
\begin{align}\label{beta_LOS}
{{{\mathbf{\bar h}}}_{kl}} = \sqrt {\beta _{kl}^{{\text{LoS}}}} e^{j\theta_{kl}} {\left[ {1,\;{e^{j2\pi {d_{\text{H}}}\sin \left( {{\varphi _{kl}}} \right)}}, \ldots, {e^{j2\pi {d_{\text{H}}}(N - 1)\sin \left( {{\varphi _{kl}}} \right)}}} \right]^{\text{T}}},
\end{align}
and
\begin{align}\label{beta_NLOS}
{\left[ {{{\mathbf{R}}_{kl}}} \right]_{s,t}} = \frac{{\beta _{kl}^{{\text{NLoS}}}}}{\bar M}\sum\limits_{m = 1}^{\bar M} {{e^{j\pi (s - t)\sin \left( {{\varphi _{kl,m}}} \right) - \frac{{\sigma _\varphi ^2}}{2}{{\left( {\pi (s - t)\cos \left( {{\varphi _{kl,m}}} \right)} \right)}^2}}}} ,
\end{align}
where ${\beta _{kl}^{{\text{LoS}}}}$ and ${{\beta _{kl}^{{\text{NLoS}}}}}$ denote the large-scale fading coefficients of LoS and NLoS components, respectively. Besides, the phase-shift is ${\theta _{kl}} \sim \mathcal{U}\left[ -\pi,+\pi \right]$, due to the sampling distance during HST driving is much greater than half-wavelength in our simulation. $\bar M$ is number of scattering clusters, and ${d_{\text{H}}} \leqslant 0.5$ is the antenna spacing parameter (in fractions of the wavelength).
In addition, ${\varphi _{kl,m}} \sim \mathcal{U}\left[ {{\varphi _{kl}} - {{30}^\text{o} },{\varphi _{kl}} + {{30}^\text{o} }} \right]$ is the nominal AoA for the $m$ cluster. The multipath components of a cluster have Gaussian distributed AoAs, distributed around the nominal AoA with the angular standard deviation (ASD). Moreover, the Rician factor is defined as $\bar K_{kl} \in \left[ { - 10:30} \right]$ dB. Then, we can compute the large-scale fading parameters for the LoS and NLoS paths in \eqref{beta_LOS} and \eqref{beta_NLOS} as $\beta _{kl}^{{\text{LoS}}} = \frac{1}{N}{\left\| {{{{\mathbf{\bar h}}}_{kl}}} \right\|^2} = \sqrt {{\bar K_{kl}}/\left( {{\bar K_{kl}} + 1} \right)} {\zeta _{kl}}$ and $\beta _{kl}^{{\text{NLoS}}} = \frac{1}{N}{\text{tr}}\left( {{{\mathbf{R}}_{kl}}} \right) = \sqrt {1/\left( {{\bar K_{kl}} + 1} \right)} {\zeta _{kl}}$, respectively.

\subsection{Results and Discussions}

\begin{figure}[t]
\begin{minipage}[t]{0.48\linewidth}	
\centering
\includegraphics[scale=0.55]{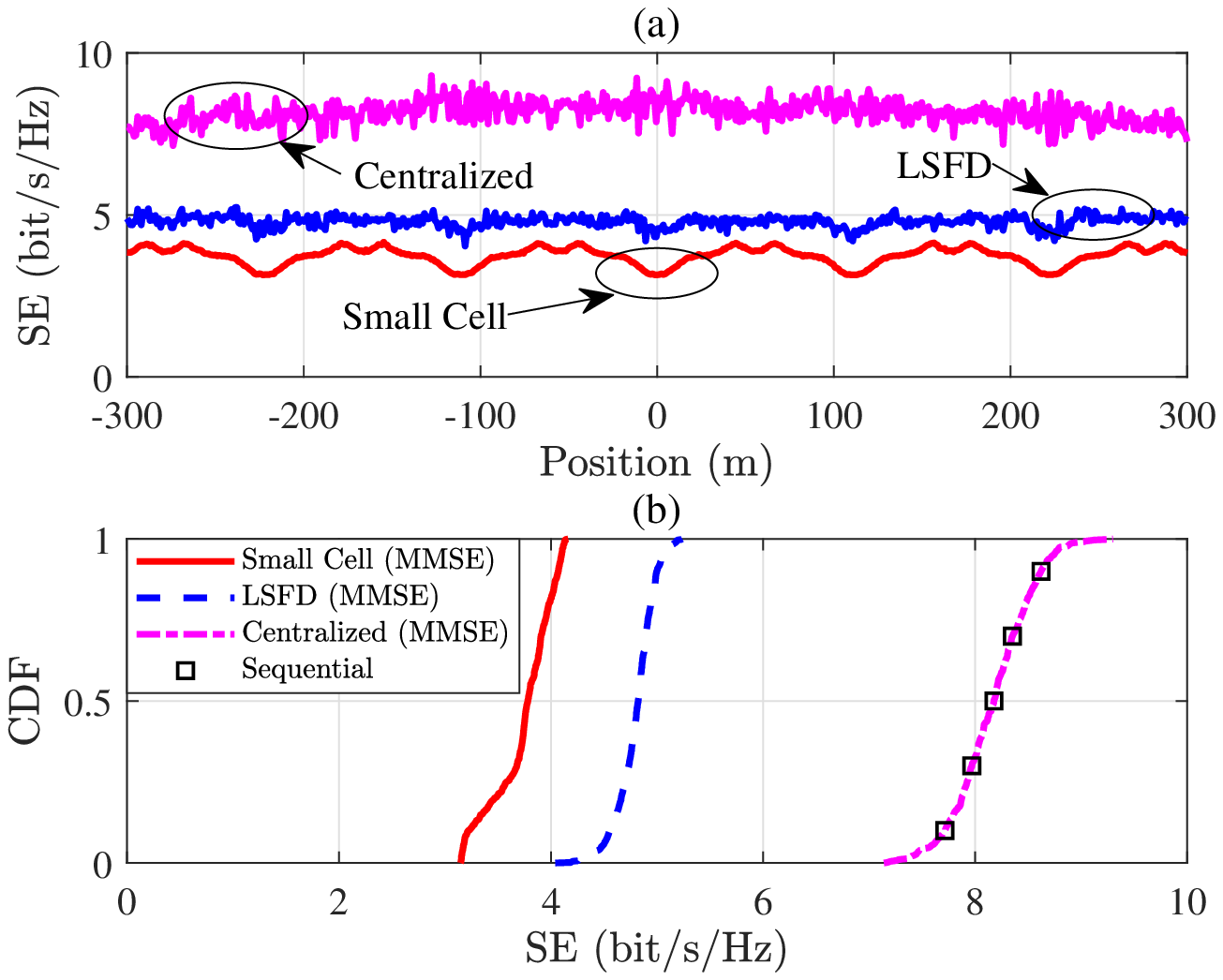}
\caption{(a) SE against the position of HST in CF and small cell massive MIMO-OFDM systems with MMSE combining. (b) CDF of per-position SE of HST in CF and small cell massive MIMO-OFDM systems with MMSE combining. ($L=10$, $K=8$, $N=4$, $v=300$ km/h, $d_\text{ve}=50$ m, $\bar K = 20$ dB, $\text{ASD}=30^{\text{o}}$)} \vspace{-4mm}
\label{fig_MMSE}
\end{minipage}
\hfill
\begin{minipage}[t]{0.48\linewidth}
\centering
\includegraphics[scale=0.55]{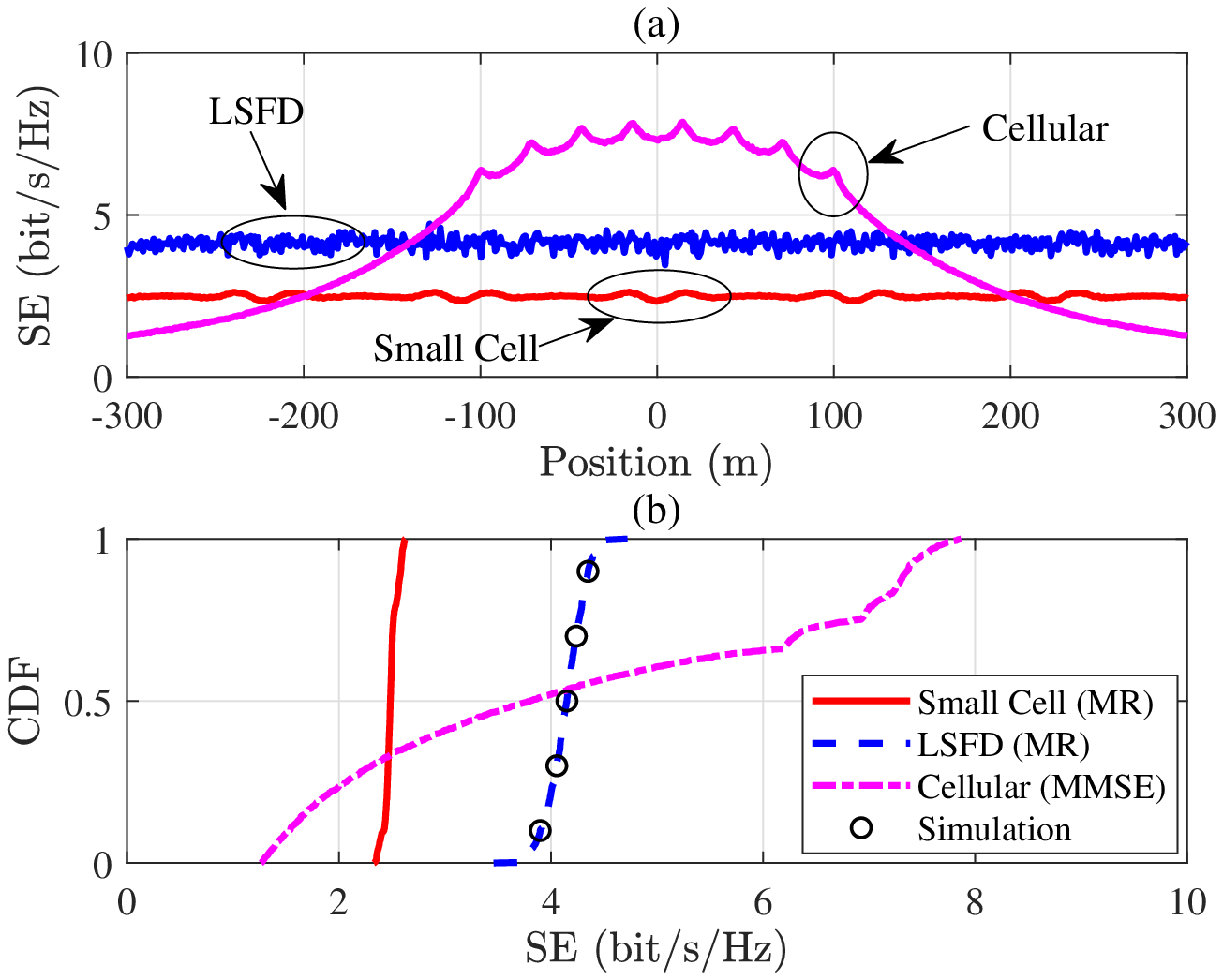}
\caption{(a) SE against the position of HST in CF, small cell systems with MR, and cellular systems with MMSE. (b) CDF of per-position SE for HST with CF, small cell systems with MR, and cellular systems with MMSE. ($L=10$, $K=8$, $N=4$, $v=300$ km/h, $d_\text{ve}=50$ m, $\bar K = 20$ dB, $\text{ASD}=30^{\text{o}}$, $L_\text{bs}=LN$)} \vspace{-4mm}
\label{fig_MR}
\end{minipage}
\end{figure}

Fig.~\ref{fig_MMSE} compares the SE of HST for CF massive MIMO-OFDM with MMSE combining under centralized processing and local processing, respectively. In addition, small cell systems are considered as a benchmark. It is found that, utilizing MMSE combining, CF massive MIMO-OFDM with centralized processing has a two-fold gain than small cell systems in HST communications. The reason is that centralized processing can use the CSI of the whole network to calculate the optimal MMSE combining vector to effectively decrease interference. In addition, by utilizing LSFD cooperation to reduce the ICI and UI effect, CF massive MIMO-OFDM with local processing achieves 28\% SE gain than the small cell. We also find that, with the change of HST position, SE performance has violent fluctuation, which is caused by phase-shift, large- and small-scale fading. In Fig.~\ref{fig_MMSE} (b), it is clear that CF massive MIMO-OFDM systems provide uniform SE performance for HST communications. Moreover, sequential linear processing proposed in \cite{shaik2020mmse} achieves the same optimal performance as centralized MMSE but with lower fronthaul requirements.

Fig.~\ref{fig_MR} shows the SE against the position of HST in the CF and small cell massive MIMO-OFDM systems with MR combining, respectively. In addition, cellular systems with MMSE combining are also considered for comparison. It is found that the SE performance of CF with LSFD cooperation is 67\% larger than the small cell in considered HST communications. The reason is that LSFD cooperation weights can effectively reduce interference from other subcarriers and TAs. Moreover, it is clear that a great SE gain can be achieved at the position near the BS, while it has poor SE performance in far positions. However, the peak SE of cellular is also lower than the SE achieved with the centralized CF with MMSE combining in Fig.~\ref{fig_MMSE}. Therefore, the centralized MMSE provides the highest SE, and is the optimal combining scheme in the considered system model. Note that the fluctuation of the SE for small cell and cellular is only caused by large- and small-scale fading and is not affected by phase-shift. From Fig.~\ref{fig_MR} (b), we can obtain that the difference between the maximum and minimum SE value in cellular is 6.6 bit/s/Hz, but the ones in LSFD and small cell are 1.2 bit/s/Hz and 0.3 bit/s/Hz, respectively. Therefore, compared with small cell and cellular systems, the CF system with LSFD cooperation can provide both large and uniform SE performance in HST communications.

\begin{figure}[t]
\begin{minipage}[t]{0.48\linewidth}	
\centering
\includegraphics[scale=0.55]{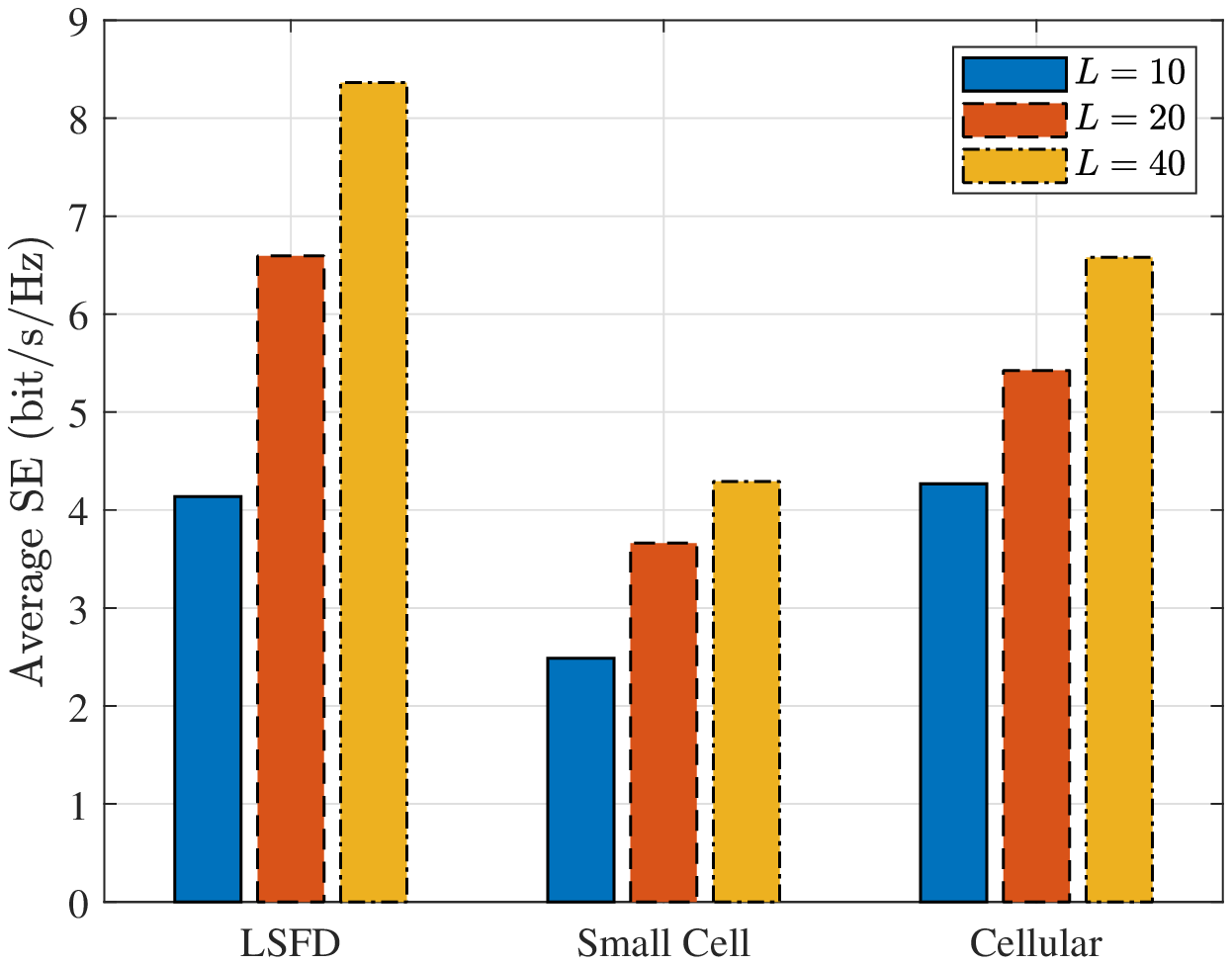}
\caption{Average SE for CF, small cell massive MIMO-OFDM systems with MR combining and cellular systems with MMSE combining ($K=8$, $N=4$, $v=300$ km/h, $d_\text{ve}=50$ m, $\bar K = 20$ dB, $\text{ASD}=30^{\text{o}}$, $L_\text{bs}=LN$).} \vspace{-4mm}
\label{fig_CF_SC_Cellular}
\end{minipage}
\hfill
\begin{minipage}[t]{0.48\linewidth}
\centering
\includegraphics[scale=0.55]{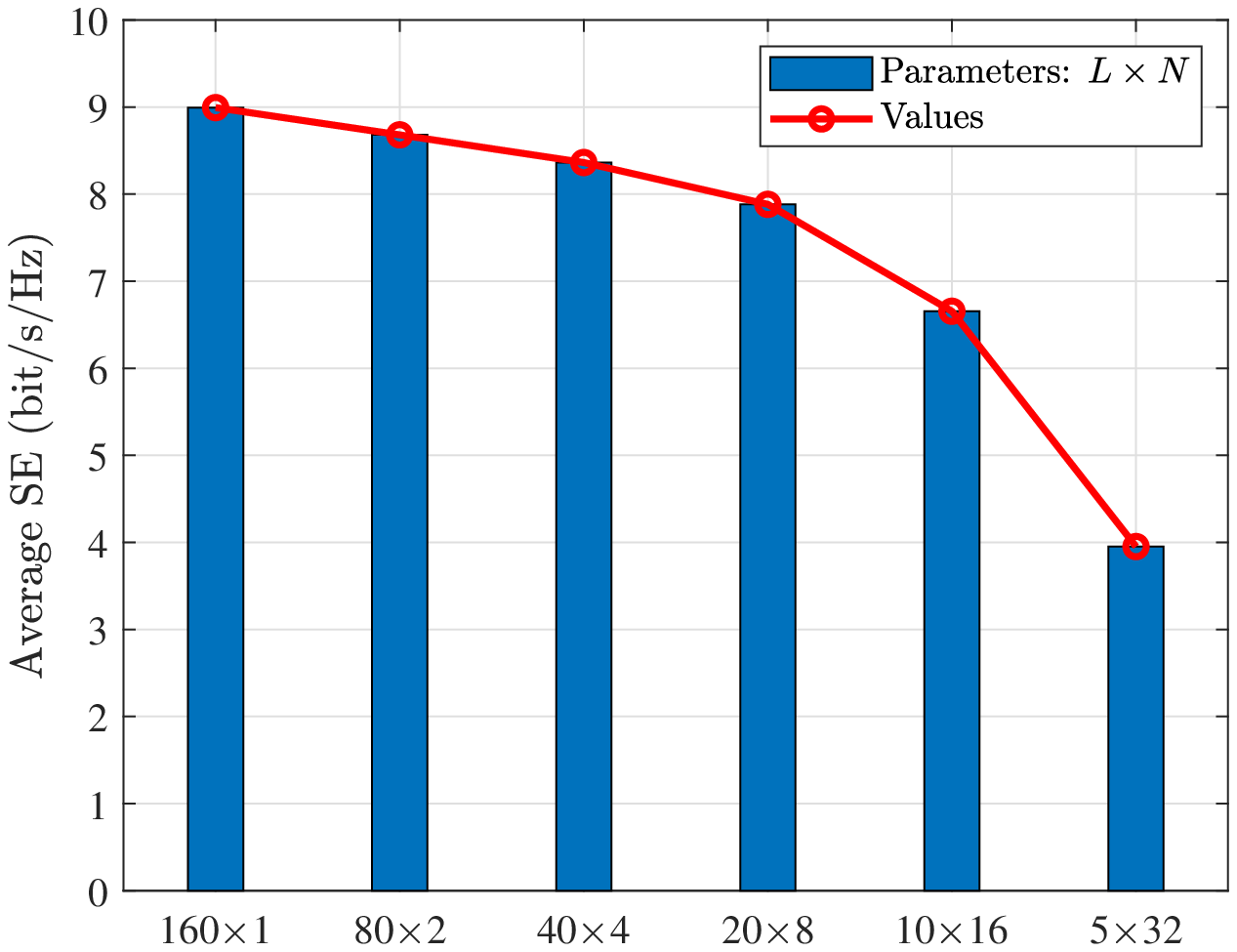}
\caption{Average SE against different deployment modes of fixed number of antennas in CF massive MIMO-OFDM systems with LSFD cooperation ($K=8$, $v=300$ km/h, $d_\text{ve} = 50$ m, $\bar K = 20$ dB, $\text{ASD}=30^{\text{o}}$).} \vspace{-4mm}
\label{LN}
\end{minipage}
\end{figure}

Fig.~\ref{fig_CF_SC_Cellular} shows the average SE for CF, small cell and cellular massive MIMO-OFDM systems with different numbers of APs, respectively. Meanwhile, CF and small cell systems use MR combining, and cellular systems utilize MMSE combining. It is found that the CF with LSFD cooperation can achieve the largest average SE among all considered systems. Moreover, as the number of APs increases from $L = 10$ to $L = 40$, the CF with LSFD cooperation can achieve two-fold SE performance gain, but not small cell and cellular systems. The reason is that LSFD cannot only obtain macro-diversity gain but also reduce interference. Therefore, we treat CF with LSFD cooperation as the best choice in HST communications, and the following analysis is based on LSFD. It is worth noting that 40 APs per kilometer and 100 MHz bandwidth are needed to achieve the goal that 800 Mbps per car in Remark \ref{goal}.

Fig.~\ref{LN} shows the average SE against different deployment modes of a fixed number of antennas in CF massive MIMO-OFDM systems with LSFD cooperation. It is found that, when there are $LN$ antennas in total per 1000 m track, CF setup with $LN$ single-antenna APs is the optimal way to distribute them along the track. The reason is that more APs can take advantage of greater spatial degrees of freedom to provide ubiquitous coverage and make the APs with small AOA always existing to reduce the DFO effect. However, more APs also increase the number of fronthaul links, resulting in complex signal processing. Therefore, a small number of antennas per AP can be selected to reduce the number of fronthaul links with little performance loss.

\begin{figure}[t]
\begin{minipage}[t]{0.48\linewidth}	
\centering
\includegraphics[scale=0.55]{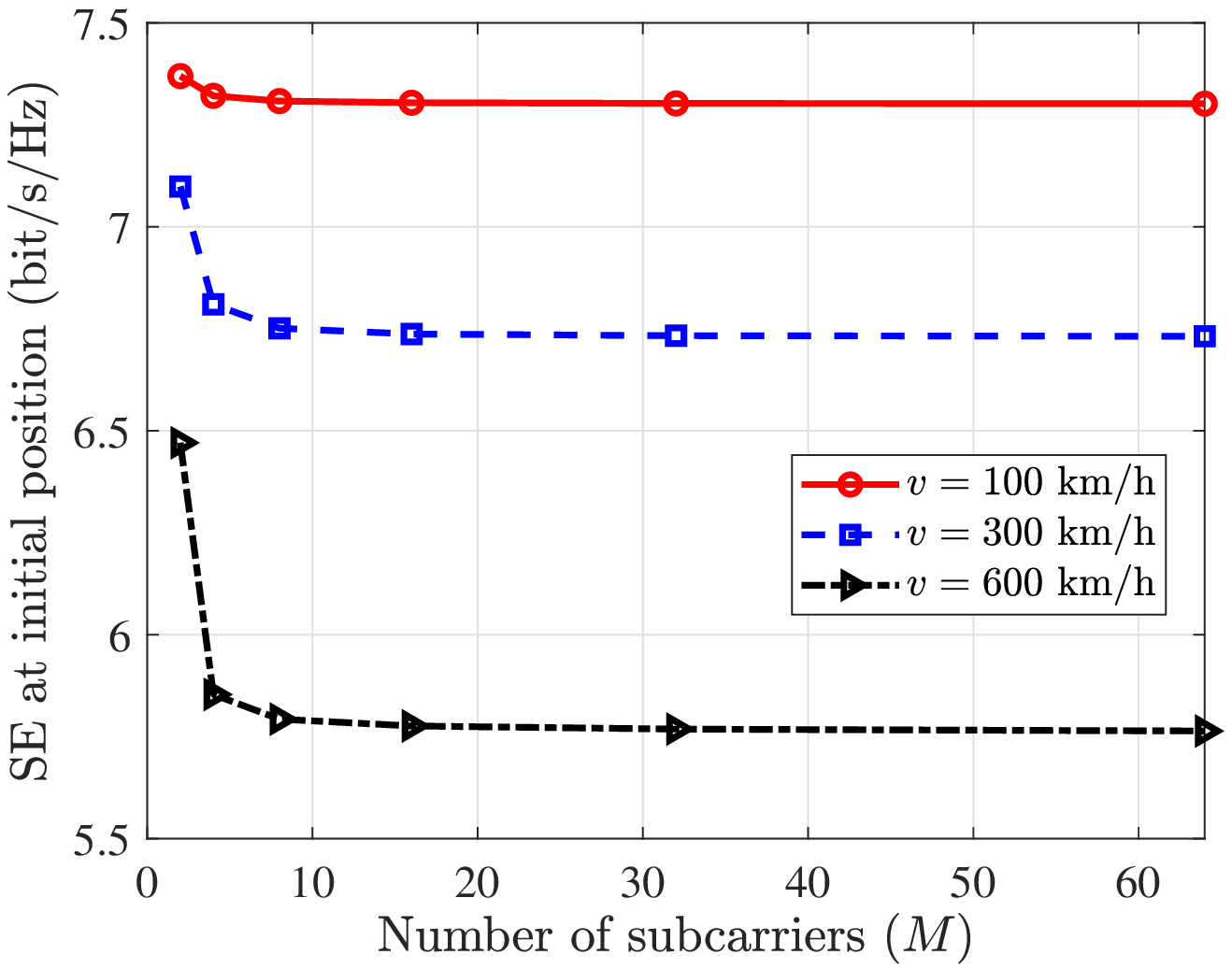}
\caption{SE against the number of subcarriers in CF massive MIMO-OFDM systems with LSFD cooperation as the train is at initial position ($L=20$, $K=8$, $N=4$, $\bar K = 20$ dB, $\text{ASD}=30^{\text{o}}$).} \vspace{-4mm}
\label{fig_SE_M}
\end{minipage}
\hfill
\begin{minipage}[t]{0.48\linewidth}
\centering
\includegraphics[scale=0.55]{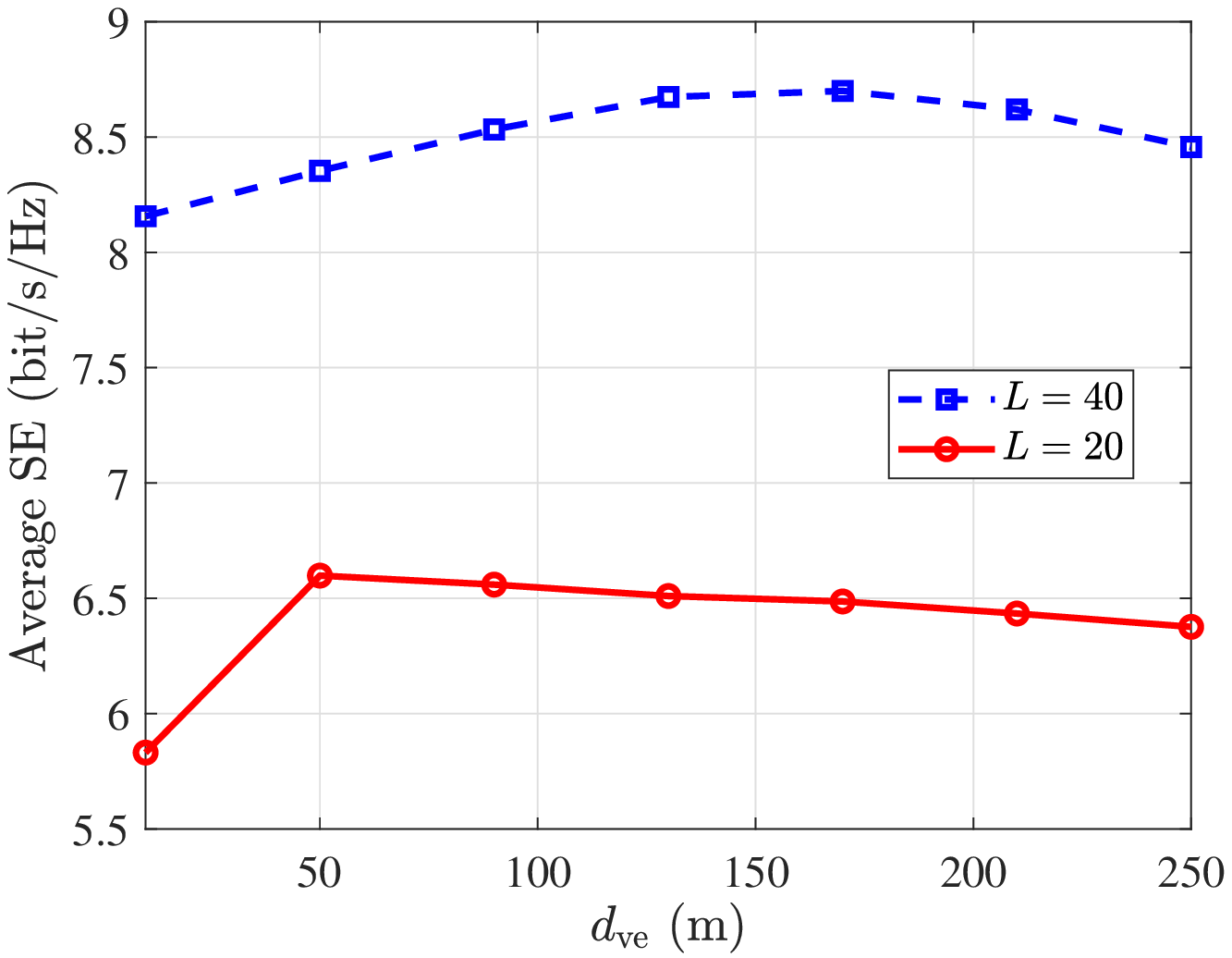}
\caption{Average SE against the distance between the rail track and APs in CF massive MIMO-OFDM systems with LSFD cooperation ($K=8$, $N=4$, $v=300$ km/h, $\bar K = 20$ dB, $\text{ASD}=30^{\text{o}}$).} \vspace{-4mm}
\label{fig_dve}
\end{minipage}
\end{figure}

Fig.~\ref{fig_SE_M} shows the SE against the number of subcarriers in CF massive MIMO-OFDM systems with LSFD cooperation as the train is at the initial position. It is clear that, with the increase of speed, SE performance is more vulnerable to the number of subcarriers. However, when the number of subcarriers exceeds 8, they all tend to be stable. That is because the ICI only occurs between adjacent subcarriers, and it also is the reason why we use one $M=8$ independent sub OFDM transmission to give representative results with good approximation.

Fig.~\ref{fig_dve} shows the average SE against the distance between the rail track and APs in the considered system with different numbers of APs. It is clear that the average SE first increases and then decreases with the increase of the distance between the rail track and APs. There is an optimal $d_\text{ve}$ to achieves the maximum average SE. The reason is that the DFO effect decreases with the distance between the rail track and APs, but the path loss increases quickly. In addition, having a larger number of APs can increase the optimal $d_\text{ve}$ at the point where we obtain maximum average SE. Moreover, as $d_\text{ve}$ decreases from 50 m to 10 m, the average SE loss for $L\!=\!20$ is larger than the case $L\!=\!40$. The reason is that a larger number of APs can reduce the DFO effect to allow APs to be closer to the TA without losing too much SE performance.

\begin{figure}[t]
\begin{minipage}[t]{0.48\linewidth}	
\centering
\includegraphics[scale=0.55]{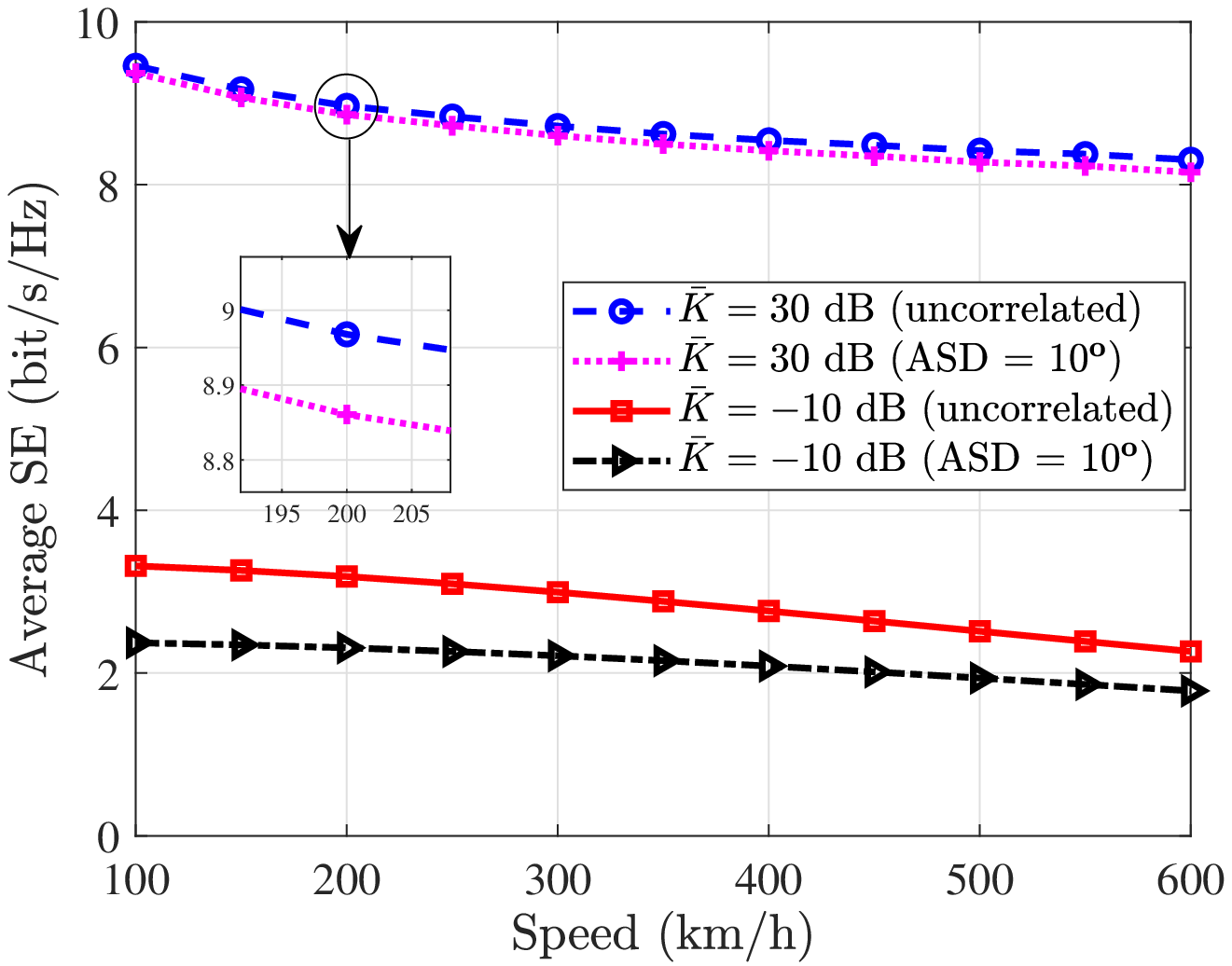}
\caption{Average SE against different speed of HST in CF massive MIMO-OFDM systems with LSFD cooperation ($L=20$, $K=8$, $N=4$, $d_\text{ve}=50$ m).} \vspace{-4mm}
\label{fig_factor}
\end{minipage}
\hfill
\begin{minipage}[t]{0.48\linewidth}
\centering
\includegraphics[scale=0.55]{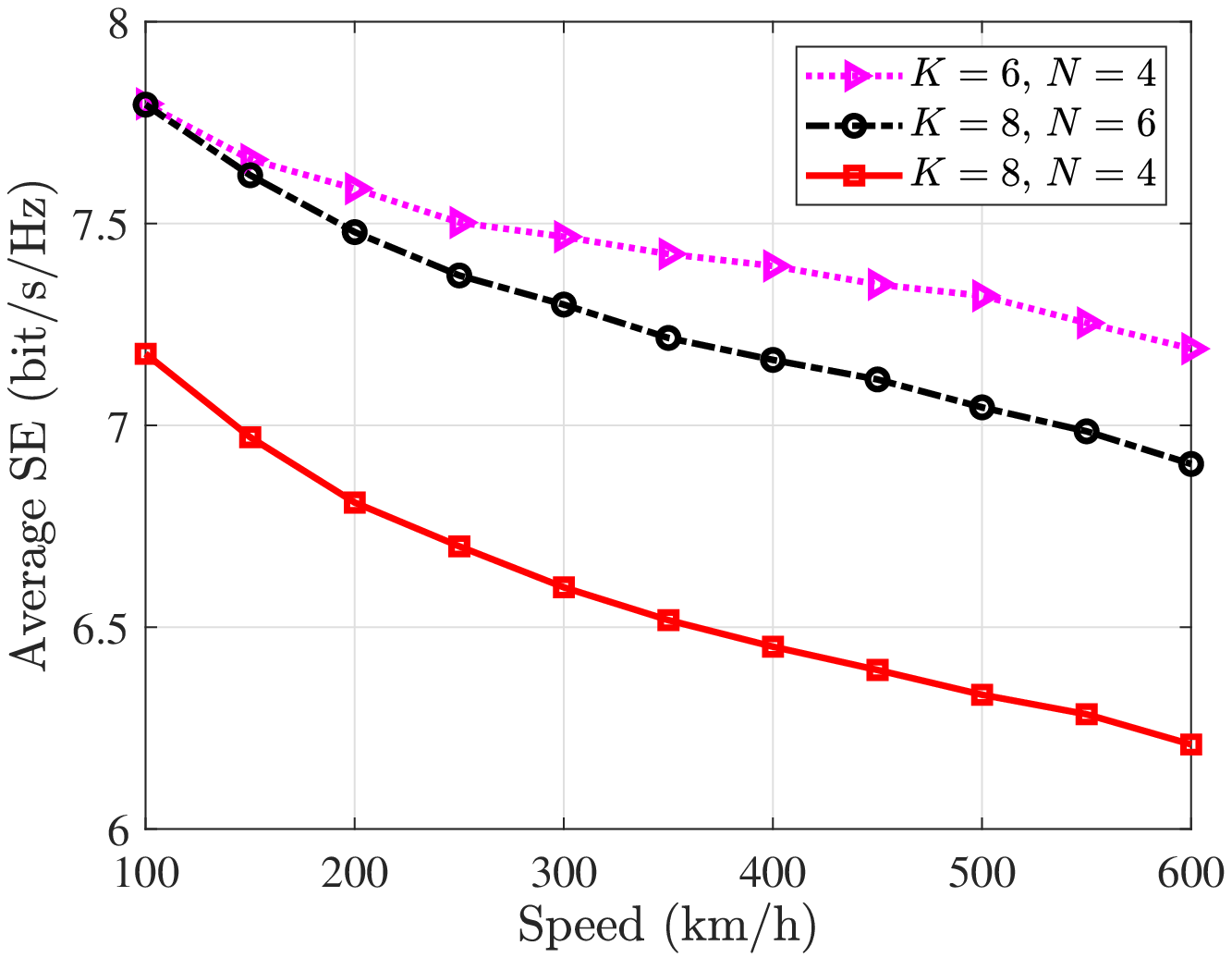}
\caption{Average SE against different speed of HST in CF massive MIMO-OFDM systems with LSFD cooperation ($L = 20$, $d_\text{ve} = 50$ m, $\bar K = 20$ dB, $\text{ASD}=30^{\text{o}}$).} \vspace{-4mm}
\label{fig_KNd}
\end{minipage}
\end{figure}

The average SE against the different speeds of HST in CF massive MIMO-OFDM systems with LSFD cooperation is shown in Fig.~\ref{fig_factor}. It is clear that the average SE decreases with the increase of the speed of the HST, due to the DFO effect. In addition, more LoS components are beneficial to the average SE, but it is more susceptible to speed. For a example, when the speed increases from 100 km/h to 600 km/h under correlated channel, the average SE for $\bar K = -10$ dB (NLoS component is main) has a 0.6 bit/s/Hz loss, but the average SE for $\bar K = 30$ dB (LoS component is main) has a 1.2 bit/s/Hz loss. Moreover, we found that uncorrelated channels can improve system performance, but the gain gradually decreases with the Rician factor increases.

Fig.~\ref{fig_KNd} compares the average SE of considered systems with different numbers of TAs $K$ and antennas per AP $N$.
It is clear that decreasing the number of TAs and increasing the number of antennas per AP can improve the SE performance. In addition, they also are available ways to reduce the influence of DFO. For example, while the speed of HST increases from 100 km/h to 600 km/h, the average SE loss of $K=8$, $N=4$ is 1.0 bit/s/Hz, but the case of $K=6$, $N=4$ is 0.8 bit/s/Hz and the case of $K=8$, $N=6$ both are 0.6 bit/s/Hz.

\begin{figure}[t]
\begin{minipage}[t]{0.48\linewidth}	
\centering
\includegraphics[scale=0.55]{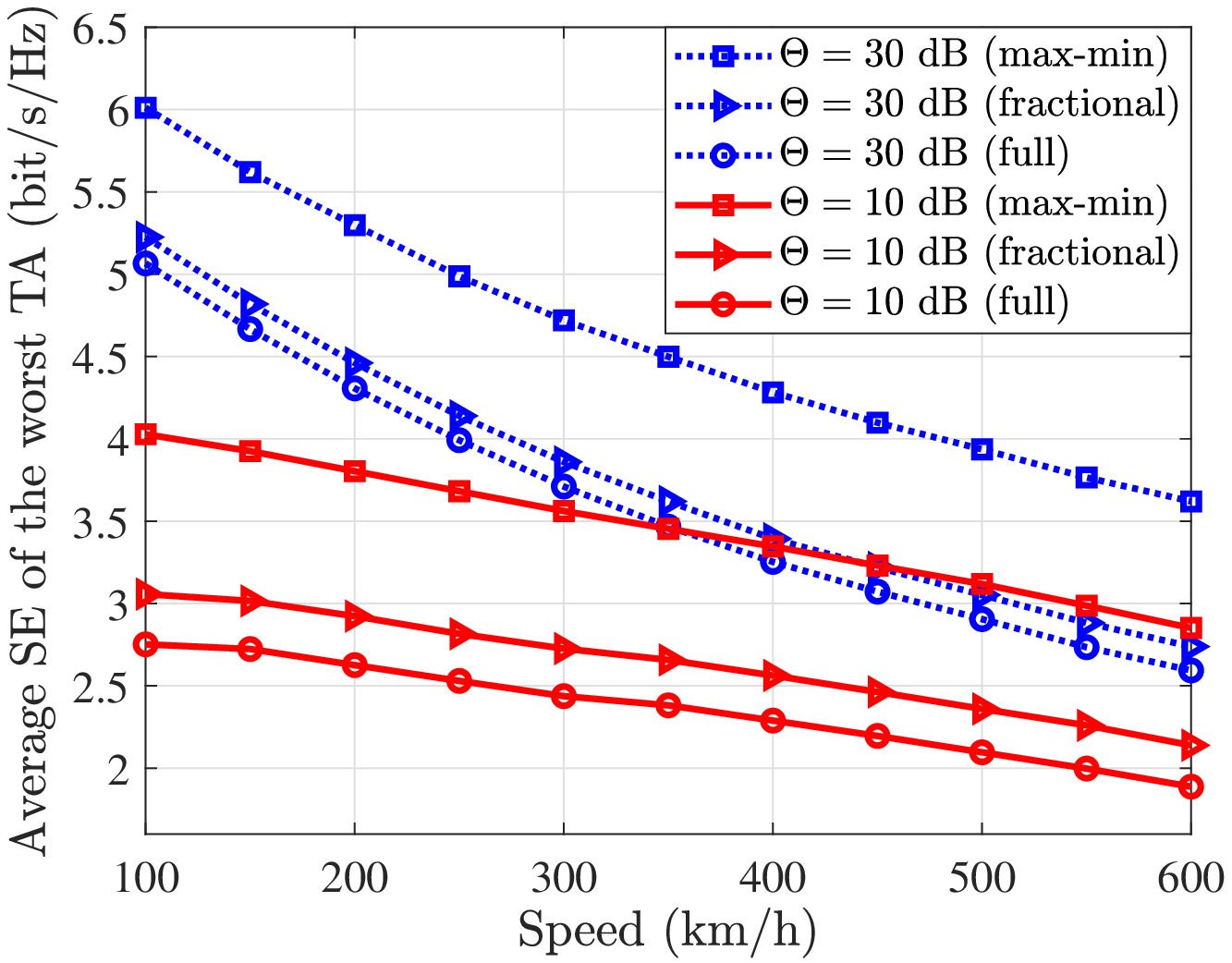}
\caption{Average SE of the worst TA against different speed of HST in TA-centric CF massive MIMO-OFDM systems with LSFD cooperation ($L=20$, $K=8$, $N=4$, $d_\text{ve}=20$ m, $\bar K = 20$ dB, $\text{ASD}=30^{\text{o}}$).}\vspace{-4mm}
\label{fig_threshold}
\end{minipage}
\hfill
\begin{minipage}[t]{0.48\linewidth}
\centering
\includegraphics[scale=0.55]{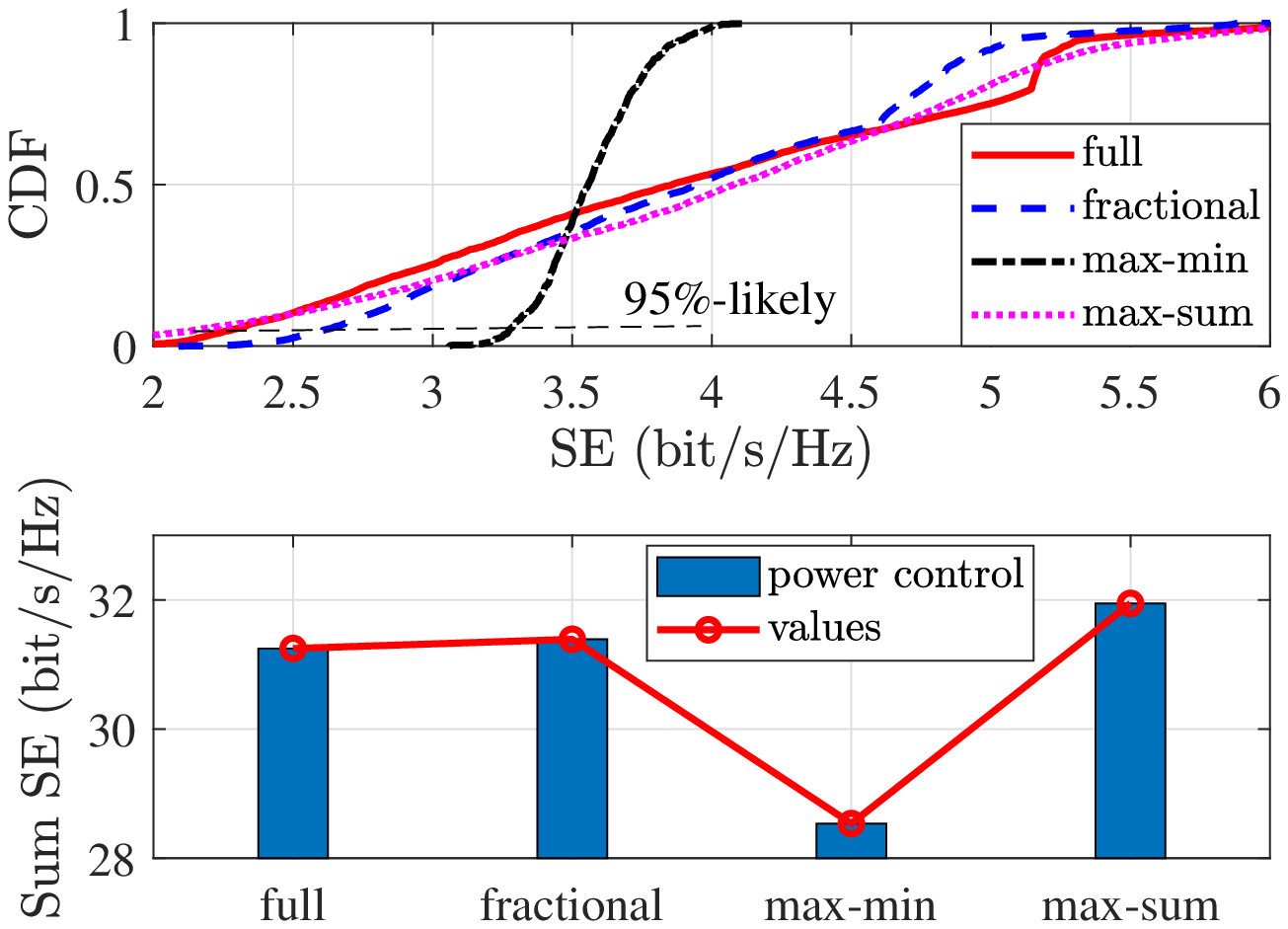}
\caption{CDF of per-position SE and sum SE against different power control schemes in TA-centric CF massive MIMO-OFDM systems with LSFD cooperation ($L=20$, $K=8$, $v=300$ km/h, $\Theta = 10$ dB, $N=4$, $d_\text{ve}=20$ m, $\bar K = 20$ dB, $\text{ASD}=30^{\text{o}}$).} \vspace{-4mm}
\label{MaxMinSumSE}
\end{minipage}
\end{figure}

The average SE of the worst TA against the different speeds of HST in TA-centric CF massive MIMO-OFDM systems with LSFD cooperation is illustrated in Fig.~\ref{fig_threshold}. The larger threshold $\Theta$ indicates that one TA is served by more nearby APs. It is clear that the average SE of the worst TA increases with the increasing of threshold, but the gain gradually reduces with the speed of HST increases. The reason is that a larger threshold $\Theta$ means more links affected by DFO are added, especially in high-speed scenarios.
It is also found that the fractional power control can efficiently improve the SE performance of the worst TA, but the gain reduces as $\Theta$ increases. Because the increase of access APs makes the channel condition of each TA tends to be consistent, which is not conducive to the fractional power control based on large-scale information.
Besides, whether it is high-speed or low-speed in Fig.~\ref{fig_threshold}, the max-min power control can bring huge SE gain to the worst TA. This is expected since the performance of the most unfortunate TA in the entire network is maximized.
However, for achieving a high level of fairness, it leads to the degradation of the sum SE performance as illustrated in Fig.~\ref{MaxMinSumSE}. The max-sum power control will result in a CDF curve that is almost entirely to the right of the competing schemes, which can achieve the largest sum SE of the considered system. Therefore, in different DFO scenarios, we can flexibly select the above three power control schemes according to the requirements of the system for complexity, fairness and total performance.

\section{Conclusions}\label{se:CON}

In this paper, considering both LoS and NLoS propagation components, we investigate the uplink SE of CF massive MIMO-OFDM systems for HST communications with large ICI caused by DFO.
Although fully centralized processing with MMSE combining achieves high-performance gain, it has large computational complexity and fronthaul requirements. Utilizing low fronthaul requirement local processing and low complexity MR combining in HST communications, we derive the closed-form expressions for uplink SE of CF massive MIMO-OFDM systems with both MF and LSFD receiver cooperation and in order to quantify the DFO effect.
In addition, HST communications with small cell and cellular massive MIMO-OFDM systems are analyzed for comparison.
It is important that the DFO effect degrades the performance of the considered systems, but the CF systems are less affected by small cell and cellular systems in HST scenarios.
In addition, increasing the number of APs significantly improves the average SE and leads to a larger distance between the rail track and APs is preferred for the optimal operating point of SE.
We also find that, compared with NLoS components, LoS components are more affected by DFO. Moreover, increasing the number of antennas and decreasing the number of TAs both can achieve large performance gains.
Furthermore, for practical application in HST scenarios, we propose a TA-centric CF-HST communication system with dynamic cooperation clustering for each TA. Finally, practical fractional power control is used in the TA-centric CF-HST system to improve the SE performance of the worst TA, and the max-min and max-sum SE power control schemes are applied for preliminary DFO interference cancellation.
In future work, we will investigate the DFO reduction in HST communications. Besides, multi-antenna TAs, non-orthogonal pilots, unknown phase-shift of the LoS link and fronthaul requirements need to be considered.

\begin{appendices}
\section{Proof of Lemma 1}

Using the inverse DFT (IDFT), the OFDM transmitted symbol is given by
\begin{align}
x\left( n \right) = \frac{1}{M}\sum\limits_{m = 0}^{M-1} {x\left[ m \right]\exp \left( {j2\pi \frac{m}{M}n} \right)} , n = 1, \dots, M-1 ,
\end{align}
where $x\left[ m \right]$ is the discrete baseband symbols on each subcarrier, $n$ is the $n$th sampling point with sampling interval $T_s/M$.
Due to the DFO, the received baseband signal is given by
\begin{align}
y\left( n \right) = x\left( n \right)\exp \left( {j2\pi \frac{n}{M}\omega \sin \left( \varphi  \right)} \right) + w\left( n \right).
\end{align}
After DFT processing the received signal is expressed as
\begin{align}
  y\left[ s \right] &= \sum\limits_{n = 0}^{M - 1} {y\left( n \right)\exp \left( { - j2\pi \frac{n}{M}s} \right) + w\left[ s \right]}  \notag \\
   &= \sum\limits_{n = 0}^{M - 1} {\frac{1}{M}\sum\limits_{m = 0}^{M - 1} {x\left[ m \right]\exp \left( {j2\pi \frac{n}{M}m} \right)} \exp \left( {j2\pi \frac{n}{M}\omega \sin \left( \varphi  \right)} \right)\exp \left( { - j2\pi \frac{n}{M}s} \right) + w\left[ s \right]}  \notag \\
   & = \sum\limits_{m = 0}^{M - 1} {x\left[ m \right]\underbrace {\frac{1}{M}\sum\limits_{n = 0}^{M - 1} {\exp \left( {j2\pi \frac{n}{M}\left( {m - s + \omega \sin \left( \varphi  \right)} \right)} \right)} }_I + w\left[ s \right]} , s = 1, \dots, M-1 ,
\end{align}
where $w\left[ s \right]$ is the DFT of the noise $w\left( n \right)$.
Utilizing the identity $ \sum\nolimits_{n = 0}^{M - 1} {{a^n} = \left( {1 - {a^M}} \right)/\left( {1 - a} \right)} $ and the trigonometric identity $\sin \left( x \right) = \left( {{e^{jx}} - {e^{ - jx}}} \right)/\left( {2j} \right)$, we have the ICI coefficients between the $m$th and $k$th subcarriers as
\begin{align}
I\left[m-s\right] = \frac{{\sin \left( {\pi \left( {m - s + \omega \sin \left( \varphi  \right)} \right)} \right)}}{{M\sin \left( {\frac{\pi }{M}\left( {m - s + \omega \sin \left( \varphi  \right)} \right)} \right)}}\exp \left( {j\pi \left( {1 - \frac{1}{{\;M}}} \right)\left( {m - s + \omega \sin \left( \varphi  \right)} \right)} \right).
\end{align}
For the LoS path with fixed $\varphi_{il}$, we obtain the ICI coefficients of LoS component as \eqref{I_kl}. For the $N_\text{path}$ NLoS paths with $\varphi  \sim U\left[ { - \pi ,\pi } \right]$, where having the feature that the AOAs and amplitudes of each NLOS paths is i.i.d. and $N_\text{path}$ is sufficiently large. Following the same steps as the proof of \cite[Lemma 1]{7921554}, we derive the statistical ICI coefficients of NLoS components as \eqref{NLOSD} to finish the proof.

\section{Proof of Theorem 3}

Using the use-and-then-forget (UatF) capacity bound in \cite{bjornson2017massive}, we obtain
\begin{align}
{\text{S}}{{\text{E}}_k}\left[ s \right] = {\log _2}\left( {1 + {\text{SIN}}{{\text{R}}_k}\left[ s \right]} \right),
\end{align}
where ${\text{SIN}}{{\text{R}}_k}\left[ s \right]$ is given as
\begin{align}
{\frac{{\mathbb{E}\left\{ {{{\left| {\text{D}{{\text{S}}_k}\left[ s \right]} \right|}^2}} \right\}}}{{\mathbb{E}\left\{ {{{\left| {{\text{B}}{{\text{U}}_k}\left[ s \right]} \right|}^2}} \right\} + \sum\limits_{m \ne s}^M {\mathbb{E}\left\{ {{{\left| {{\text{IC}}{{\text{I}}_k}\left[ {m - s} \right]} \right|}^2}} \right\}}  + \sum\limits_{i \ne k}^K {\sum\limits_{m = 1}^M {\mathbb{E}\left\{ {{{\left| {{\text{U}}{{\text{I}}_{ki}}\left[ {m - s} \right]} \right|}^2}} \right\}} }  + \mathbb{E}\left\{ {{{\left| {{\text{N}}{{\text{S}}_k}\left[ s \right]} \right|}^2}} \right\}}}}.
\end{align}
We will compute every term of ${\text{SIN}}{{\text{R}}_k}\left[ s \right]$ to obtain \eqref{LSFD_MR}.

\emph{1) Compute ${\mathbb{E}\left\{ {{{\left| {{\mathrm{DS}}_k}\left[ s \right] \right|}^2}} \right\}}$:}
Based on the properties of MMSE estimation, ${\mathbf{\hat g}}_{kl}$ and ${{{{\mathbf{\tilde g}}}_{kl}}}$ (or ${{{{\mathbf{\tilde h}}}_{kl}}}$) are independent, we then have
\begin{align}\label{DS}
&\!\!\!\mathbb{E}\left\{ {{\mathbf{\hat g}}_{kl}^{\text{H}}\left( {{I_{kl}}\left[ 0 \right]{{{\mathbf{\bar h}}}_{kl}} + {I_{D}}\left[ 0 \right]{{\mathbf{h}}_{kl}}} \right)} \right\} = \mathbb{E}\left\{ {{{\left( {{{{\mathbf{\bar h}}}_{kl}} + {{{\mathbf{\hat h}}}_{kl}}} \right)}^{\text{H}}}\left( {{I_{kl}}\left[ 0 \right]{{{\mathbf{\bar h}}}_{kl}} + {I_{D}}\left[ 0 \right]\left( {{{{\mathbf{\hat h}}}_{kl}} + {{{\mathbf{\tilde h}}}_{kl}}} \right)} \right)} \right\} \notag \\
   &\;\;\;\;\;\;\;\;\;\;\;\;\;= {I_{kl}}\left[ 0 \right]\mathbb{E}\left\{ {{\mathbf{\bar h}}_{kl}^{\text{H}}{{{\mathbf{\bar h}}}_{kl}}} \right\} + {I_{D}}\left[ 0 \right]\mathbb{E}\left\{ {{\mathbf{\hat h}}_{kl}^{\text{H}}{{{\mathbf{\hat h}}}_{kl}}} \right\} = {I_{kl}}\left[ 0 \right]{\mathbf{\bar h}}_{kl}^{\text{H}}{{{\mathbf{\bar h}}}_{kl}} + {I_{D}}\left[ 0 \right]{\text{tr}}\left( {{{\mathbf{Q}}_{kl}}} \right).
\end{align}
Furthermore, we can obtain
\begin{align}
\mathbb{E}\left\{ {{{\left| {\text{D}{{\text{S}}_k}\left[ s \right]} \right|}^2}} \right\} = {\left| {\sum\limits_{l = 1}^L {\sqrt {{p_k}} a_{kl}^*\left[ s \right]\left( {{I_{kl}}\left[ 0 \right]{\mathbf{\bar h}}_{kl}^{\text{H}}{{{\mathbf{\bar h}}}_{kl}} + {I_{D}}\left[ 0 \right]{\text{tr}}\left( {{{\mathbf{Q}}_{kl}}} \right)} \right)} } \right|^2}.
\end{align}

\emph{2) Compute ${\mathbb{E}\left\{ {{{\left| {{\mathrm{BU}}_k}\left[ s \right] \right|}^2}} \right\}}$:} With the help of \cite[Eq. (76)]{9453784}, we have
\begin{align}\label{BU}
  &\mathbb{E}\left\{ {{{\left| {{\text{B}}{{\text{U}}_k}\left[ s \right]} \right|}^2}} \right\} = {p_k}\sum\limits_{l = 1}^L {{{\left| {a_{kl}^*\left[ s \right]} \right|}^2}\mathbb{E}\left\{ {{{\left| {{\mathbf{\hat g}}_{kl}^{\text{H}}\left( {{I_{kl}}\left[ 0 \right]{{{\mathbf{\bar h}}}_{kl}} + {I_{D}}\left[ 0 \right]{{\mathbf{h}}_{kl}}} \right)} \right|}^2}} \right\}}  \notag \\
   &+ {p_k}\!\sum\limits_{l = 1}^L \!{\sum\limits_{n \ne l}^L {{a_{kl}}\left[ s \right]a_{kn}^*\left[ s \right]\mathbb{E}\!\left\{ \! {{{\left( {{\mathbf{\hat g}}_{kl}^{\text{H}}\left( {{I_{kl}}\left[ 0 \right]{{{\mathbf{\bar h}}}_{kl}} \!+\! {I_{D}}\left[ 0 \right]{{\mathbf{h}}_{kl}}} \right)} \right)}^*}\left( {{\mathbf{\hat g}}_{kn}^{\text{H}}\left( {{I_{kn}}\left[ 0 \right]{{{\mathbf{\bar h}}}_{kn}} \!+\! {I_{D}}\left[ 0 \right]{{\mathbf{h}}_{kn}}} \right)} \right)} \!\right\}} }  \notag \\
   &- {\left| {\sum\limits_{l = 1}^L {\sqrt {{p_k}} a_{kl}^*\left[ s \right]\mathbb{E}\left\{ {{\mathbf{\hat g}}_{kl}^{\text{H}}\left( {{I_{kl}}\left[ 0 \right]{{{\mathbf{\bar h}}}_{kl}} + {I_{D}}\left[ 0 \right]{{\mathbf{h}}_{kl}}} \right)} \right\}} } \right|^2} .
\end{align}
We first calculate the first term of \eqref{BU} as
\begin{align}\label{a1}
  &\mathbb{E}\!\left\{\! {{{\left| {{\mathbf{\hat g}}_{kl}^{\text{H}}\left( {{I_{kl}}\left[ 0 \right]{{{\mathbf{\bar h}}}_{kl}} \!+\! {I_{D}}\left[ 0 \right]\left( {{{{\mathbf{\hat h}}}_{kl}} \!+\! {{{\mathbf{\tilde h}}}_{kl}}} \right)} \right)} \right|}^2}} \!\right\} \!=\! \mathbb{E}\!\left\{\! {{{\left| {{\mathbf{\hat g}}_{kl}^{\text{H}}\left( {{I_{kl}}\left[ 0 \right]{{{\mathbf{\bar h}}}_{kl}} \!+\! {I_{D}}\left[ 0 \right]{{{\mathbf{\hat h}}}_{kl}}} \right) \!+\! {I_{D}}\left[ 0 \right]{\mathbf{\hat g}}_{kl}^{\text{H}}{{{\mathbf{\tilde h}}}_{kl}}} \right|}^2}} \!\right\} \notag \\
   &= \mathbb{E}\left\{ {{{\left| {{\mathbf{\hat g}}_{kl}^{\text{H}}\left( {{I_{kl}}\left[ 0 \right]{{{\mathbf{\bar h}}}_{kl}} + {I_{D}}\left[ 0 \right]{{{\mathbf{\hat h}}}_{kl}}} \right)} \right|}^2}} \right\} + \mathbb{E}\left\{ {{{\left| {{I_{D}}\left[ 0 \right]{\mathbf{\hat g}}_{kl}^{\text{H}}{{{\mathbf{\tilde h}}}_{kl}}} \right|}^2}} \right\} .
\end{align}
In order to derive the closed-form expression for \eqref{a1}, note that ${\mathbf{\hat g}}_{kl}^{\text{H}} = {{{\mathbf{\bar h}}}_{kl}} + {{{\mathbf{\hat h}}}_{kl}} = {{{\mathbf{\bar h}}}_{kl}} + {\mathbf{Q}}_{kl}^{\frac{1}{2}}{\mathbf{m}}$ and ${{I_{kl}}\left[ 0 \right]{{{\mathbf{\bar h}}}_{kl}} \!+\! {I_{D}}\left[ 0 \right]{{{\mathbf{\hat h}}}_{kl}}} \!=\! {I_{kl}}\left[ 0 \right]{{{\mathbf{\bar h}}}_{kl}} \!+\! {I_{D}}\left[ 0 \right]{\mathbf{Q}}_{kl}^{\frac{1}{2}}{\mathbf{m}}$,
where ${\mathbf{m}}\sim\mathcal{C}\mathcal{N}\left( {{\mathbf{0}},{{\mathbf{I}}_N}} \right)$. Then, we can derive
\begin{align}
  \mathbb{E}\left\{ {{{\left| {{\mathbf{\hat g}}_{kl}^{\text{H}}\left( {{I_{kl}}\left[ 0 \right]{{{\mathbf{\bar h}}}_{kl}} + {I_{D}}\left[ 0 \right]{{{\mathbf{\hat h}}}_{kl}}} \right)} \right|}^2}} \right\} = \mathbb{E}\left\{ {\left| {\underbrace {{{\mathbf{m}}^{\text{H}}}{{\left( {{\mathbf{Q}}_{kl}^{\text{H}}} \right)}^{\frac{1}{2}}}{I_{D}}\left[ 0 \right]{\mathbf{Q}}_{kl}^{\frac{1}{2}}{\mathbf{m}}}_{a\left[ 0 \right]}} \right.} \right. \notag \\
  \left. {{{\left. { + \underbrace {{{\mathbf{m}}^{\text{H}}}{{\left( {{\mathbf{Q}}_{kl}^{\text{H}}} \right)}^{\frac{1}{2}}}{I_{kl}}\left[ 0 \right]{{{\mathbf{\bar h}}}_{kl}}}_{b\left[ 0 \right]} + \underbrace {{\mathbf{\bar h}}_{kl}^{\text{H}}{I_{D}}\left[ 0 \right]{\mathbf{Q}}_{kl}^{\frac{1}{2}}{\mathbf{m}}}_{c\left[ 0 \right]} + \underbrace {{\mathbf{\bar h}}_{kl}^{\text{H}}{I_{kl}}\left[ 0 \right]{{{\mathbf{\bar h}}}_{kl}}}_{d\left[ 0 \right]}} \right|}^2}} \right\} .
\end{align}
With the help of \cite[(67)-(70)]{9453784}, we can calculate all nonzero terms $\mathbb{E}\!\left\{ {a\!\left[ 0 \right]\!{a^*}\!\left[ 0 \right]} \right\}$, $\mathbb{E}\left\{ {b\left[ 0 \right]{b^*}\left[ 0 \right]} \right\}$, $\mathbb{E}\left\{ {c\left[ 0 \right]{c^*}\left[ 0 \right]} \right\}$, $\mathbb{E}\left\{ {d\left[ 0 \right]{d^*}\left[ 0 \right]} \right\}$, $\mathbb{E}\left\{ {a\left[ 0 \right]{d^*}\left[ 0 \right]} \right\}$ and $\mathbb{E}\left\{ {d\left[ 0 \right]{a^*}\left[ 0 \right]} \right\}$.
We then can obtain
\begin{align}\label{s1}
  \mathbb{E}&\left\{ {{{\left| {{\mathbf{\hat g}}_{kl}^{\text{H}}\left( {{I_{kl}}\left[ 0 \right]{{{\mathbf{\bar h}}}_{kl}} + {I_{D}}\left[ 0 \right]{{{\mathbf{\hat h}}}_{kl}}} \right)} \right|}^2}} \right\} = {\left( {{p_k}{\tau _p}{I_{D}}\left[ 0 \right]} \right)^2}{\left| {{\text{tr}}\left( {{{\mathbf{R}}_{kl}}{{\mathbf{\Psi }}_{kl}}{{\mathbf{R}}_{kl}}} \right)} \right|^2} + I_{kl}^2\left[ 0 \right]\left| {{\mathbf{\bar h}}_{kl}^{\text{H}}{{{\mathbf{\bar h}}}_{kl}}} \right| \notag \\
   &+ {p_k}{\tau _p}I_{D}^2\left[ 0 \right]{\text{tr}}\left( {\left( {{{\mathbf{R}}_{kl}} - {{\mathbf{C}}_{kl}}} \right){{\mathbf{R}}_{kl}}{{\mathbf{\Psi }}_{kl}}{{\mathbf{R}}_{kl}}} \right) + {p_k}{\tau _p}I_{kl}^2\left[ 0 \right]{\mathbf{\bar h}}_{kl}^{\text{H}}{{\mathbf{R}}_{kl}}{{\mathbf{\Psi }}_{kl}}{{\mathbf{R}}_{kl}}{{{\mathbf{\bar h}}}_{kl}} \notag \\
   &+ I_{D}^2\left[ 0 \right]{\mathbf{\bar h}}_{kl}^{\text{H}}\left( {{{\mathbf{R}}_{kl}} - {{\mathbf{C}}_{kl}}} \right){{{\mathbf{\bar h}}}_{kl}} + 2{p_k}{\tau _p}{I_{D}}\left[ 0 \right]{I_{kl}}\left[ 0 \right]\operatorname{Re} \left\{ {{\text{tr}}\left( {{{\mathbf{R}}_{kl}}{{\mathbf{\Psi }}_{kl}}{{\mathbf{R}}_{kl}}} \right){\mathbf{\bar h}}_{kl}^{\text{H}}{{{\mathbf{\bar h}}}_{kl}}} \right\} .
\end{align}
In addition, we also derive
\begin{align}\label{s2}
\mathbb{E}\left\{ {{{\left| {{I_{D}}\left[ 0 \right]{\mathbf{\hat g}}_{kl}^{\text{H}}{{{\mathbf{\tilde h}}}_{kl}}} \right|}^2}} \right\}{\text{ = }}I_{D}^2\left[ 0 \right]\left( {{p_k}{\tau _p}{\text{tr}}\left( {{{\mathbf{C}}_{kl}}{{\mathbf{R}}_{kl}}{{\mathbf{\Psi }}_{kl}}{{\mathbf{R}}_{kl}}} \right) + {\mathbf{\bar h}}_{kl}^{\text{H}}{{\mathbf{C}}_{kl}}{{{\mathbf{\bar h}}}_{kl}}} \right) .
\end{align}
With the help of \eqref{s1} and \eqref{s2}, we have
\begin{align}\label{BU1}
  &\mathbb{E}\left\{ {{{\left| {{\mathbf{\hat g}}_{kl}^{\text{H}}\left( {{I_{kl}}\left[ 0 \right]{{{\mathbf{\bar h}}}_{kl}} \!+\! {I_{D}}\left[ 0 \right]{{\mathbf{h}}_{kl}}} \right)} \right|}^2}} \right\} \!=\! I_D^2\left[ 0 \right]{\left| {{\text{tr}}\left( {{{\mathbf{Q}}_{kl}}} \right)} \right|^2} \!+\! I_{kl}^2\left[ 0 \right]{\left| {{\mathbf{\bar h}}_{kl}^{\text{H}}{{{\mathbf{\bar h}}}_{kl}}} \right|^2} \!+\! I_{D}^2\left[ 0 \right]{\text{tr}}\left( {{{\mathbf{R}}_{kl}}{{\mathbf{Q}}_{kl}}} \right) \notag \\
   &\;\;\;\;\;\;\;\;\;\;+ I_{kl}^2\left[ 0 \right]{\mathbf{\bar h}}_{kl}^{\text{H}}{{\mathbf{Q}}_{kl}}{{{\mathbf{\bar h}}}_{kl}} + I_{D}^2\left[ 0 \right]{\mathbf{\bar h}}_{kl}^{\text{H}}{{\mathbf{R}}_{kl}}{{{\mathbf{\bar h}}}_{kl}} + 2{I_{D}}\left[ 0 \right]{I_{kl}}\left[ 0 \right]\operatorname{Re} \left\{ {{\text{tr}}\left( {{{\mathbf{Q}}_{kl}}} \right){\mathbf{\bar h}}_{kl}^{\text{H}}{{{\mathbf{\bar h}}}_{kl}}} \right\} .
\end{align}
Submitting \eqref{DS} and \eqref{BU1} into \eqref{BU}, we obtain that
\begin{align}
  &\mathbb{E}\left\{ {{{\left| {{\text{B}}{{\text{U}}_k}\left[ s \right]} \right|}^2}} \right\} = {p_k}{\left| {\sum\limits_{l = 1}^L {a_{kl}^*\left[ s \right]} \left( {{I_{kl}}\left[ 0 \right]{\mathbf{\bar h}}_{kl}^{\text{H}}{{{\mathbf{\bar h}}}_{kl}} + {I_{D}}\left[ 0 \right]{\text{tr}}\left( {{{\mathbf{Q}}_{kl}}} \right)} \right)} \right|^2} \notag \\
   &+ {p_k}\sum\limits_{l = 1}^L {{{\left| {a_{kl}^*\left[ s \right]} \right|}^2}\left( {I_{D}^2\left[ 0 \right]{\text{tr}}\left( {{{\mathbf{R}}_{kl}}{{\mathbf{Q}}_{kl}}} \right) + I_{kl}^2\left[ 0 \right]{\mathbf{\bar h}}_{kl}^{\text{H}}{{\mathbf{Q}}_{kl}}{{{\mathbf{\bar h}}}_{kl}} + I_{D}^2\left[ 0 \right]{\mathbf{\bar h}}_{kl}^{\text{H}}{{\mathbf{R}}_{kl}}{{{\mathbf{\bar h}}}_{kl}}} \right)} .
\end{align}

\emph{3) Compute ${\mathbb{E}\left\{ {{{\left| {{\mathrm{ICI}}_k}\left[ m-s \right] \right|}^2}} \right\}}$:} Following the similar steps obtained ${\mathbb{E}\left\{ {{{\left| {{\mathrm{BU}}_k}\left[ s \right] \right|}^2}} \right\}}$, we can derive
\begin{align}
  &\mathbb{E}\left\{ {{{\left| {{\text{IC}}{{\text{I}}_k}\left[ {m - s} \right]} \right|}^2}} \right\} = \mathbb{E}\left\{ {{{\left| {\sum\limits_{l = 1}^L {\sqrt {{p_k}} a_{kl}^*\left[ s \right]{\mathbf{\hat g}}_{kl}^{\text{H}}\left( {{I_{kl}}\left[ {m - s} \right]{{{\mathbf{\bar h}}}_{kl}} + {I_{D}}\left[ {m - s} \right]{{\mathbf{h}}_{kl}}} \right)} } \right|}^2}} \right\} \notag\\
   &= {p_k}{\left| {\sum\limits_{l = 1}^L {a_{kl}^*\left[ s \right]} \left( {{I_{kl}}\left[ {m - s} \right]{\mathbf{\bar h}}_{kl}^{\text{H}}{{{\mathbf{\bar h}}}_{kl}} + {I_{D}}\left[ {m - s} \right]{\text{tr}}\left( {{{\mathbf{Q}}_{kl}}} \right)} \right)} \right|^2} \notag \\
   &+\! {p_k}\sum\limits_{l = 1}^L {{{\left| {a_{kl}^*\left[ s \right]} \right|}^2}\left( {I_{D}^2\left[ {m \!-\! s} \right]{\text{tr}}\left( {{{\mathbf{R}}_{kl}}{{\mathbf{Q}}_{kl}}} \right) \!+\! I_{kl}^2\left[ {m \!-\! s} \right]{\mathbf{\bar h}}_{kl}^{\text{H}}{{\mathbf{Q}}_{kl}}{{{\mathbf{\bar h}}}_{kl}} \!+\! I_{D}^2\left[ {m \!-\! s} \right]{\mathbf{\bar h}}_{kl}^{\text{H}}{{\mathbf{R}}_{kl}}{{{\mathbf{\bar h}}}_{kl}}} \right)}  .
\end{align}

\emph{4) Compute ${\mathbb{E}\left\{ {{{\left| {{\mathrm{UI}}_{ki}}\left[ m-s \right] \right|}^2}} \right\}}$:} Following the \eqref{BU}, expand the UI term as
\begin{align}\label{UI}
  &\mathbb{E}\!\left\{ {{{\left| {{\text{U}}{{\text{I}}_{ki}}\!\left[ {m \!-\! s} \right]} \right|}^2}} \right\} \!=\! {p_i}\sum\limits_{l = 1}^L {{{\left| {a_{kl}^*\left[ s \right]} \right|}^2}\mathbb{E}\!\left\{ {{{\left| {{\mathbf{\hat g}}_{kl}^{\text{H}}\left( {{I_{il}}\!\left[ {m \!-\! s} \right]{{{\mathbf{\bar h}}}_{il}} \!+\! {I_{D}}\!\left[ {m \!-\! s} \right]{{\mathbf{h}}_{il}}} \right)} \right|}^2}} \right\}} \!+\! {p_i}\sum\limits_{l = 1}^L {\sum\limits_{n \ne l}^L {{a_{kl}}\!\left[ s \right]} }  \notag \\
   &\times a_{kn}^*\!\left[ s \right]\mathbb{E}\!\left\{ {{{\left( {{\mathbf{\hat g}}_{kl}^{\text{H}}\left( {{I_{il}}\!\left[ {m \!-\! s} \right]{{{\mathbf{\bar h}}}_{il}} \!+\! {I_{D}}\!\left[ {m \!-\! s} \right]{{\mathbf{h}}_{il}}} \right)} \right)}^*}\left( {{\mathbf{\hat g}}_{kn}^{\text{H}}\left( {{I_{in}}\!\left[ {m \!-\! s} \right]{{{\mathbf{\bar h}}}_{in}} \!+\! {I_{D}}\!\left[ {m \!-\! s} \right]{{\mathbf{h}}_{in}}} \right)} \right)} \right\} .
\end{align}
It is similar with \eqref{BU1}, but is for different independent TAs (i.e., $i \ne k$). Therefore, we can compute the first term of \eqref{UI} as
\begin{align}\label{UI1}
  &\mathbb{E}\left\{ {{{\left| {{\mathbf{\hat g}}_{kl}^{\text{H}}\left( {{I_{il}}\left[ {m - s} \right]{{{\mathbf{\bar h}}}_{il}} + {I_{D}}\left[ {m - s} \right]{{{\mathbf{\hat h}}}_{il}}} \right)} \right|}^2}} \right\} + \mathbb{E}\left\{ {{{\left| {{I_{D}}\left[ {m - s} \right]{\mathbf{\hat g}}_{kl}^{\text{H}}{{{\mathbf{\tilde h}}}_{il}}} \right|}^2}} \right\}  \notag\\
   &= I_{D}^2\left[ {m - s} \right]{\text{tr}}\left( {{{\mathbf{Q}}_{il}}{{\mathbf{Q}}_{kl}}} \right) + I_{D}^2\left[ {m - s} \right]{\mathbf{\bar h}}_{kl}^{\text{H}}{{\mathbf{Q}}_{il}}{{{\mathbf{\bar h}}}_{kl}} + I_{il}^2\left[ {m - s} \right]{\mathbf{\bar h}}_{il}^{\text{H}}{{\mathbf{Q}}_{kl}}{{{\mathbf{\bar h}}}_{il}} \notag\\
   &+ I_{il}^2\left[ {m - s} \right]{\left| {{\mathbf{\bar h}}_{kl}^{\text{H}}{{{\mathbf{\bar h}}}_{il}}} \right|^2} + I_{D}^2\left[ {m - s} \right]\left( {{\text{tr}}\left( {{{\mathbf{C}}_{il}}{{\mathbf{Q}}_{kl}}} \right) + {\mathbf{\bar h}}_{kl}^{\text{H}}{{\mathbf{C}}_{il}}{{{\mathbf{\bar h}}}_{kl}}} \right) \notag \\
   &=\! I_{D}^2\left[ {m \!-\! s} \right]{\text{tr}}\left( {{{\mathbf{R}}_{il}}{{\mathbf{Q}}_{kl}}} \right) \!+\! I_{D}^2\left[ {m \!-\! s} \right]{\mathbf{\bar h}}_{kl}^{\text{H}}{{\mathbf{R}}_{il}}{{{\mathbf{\bar h}}}_{kl}} \!+\! I_{il}^2\left[ {m \!-\! s} \right]{\mathbf{\bar h}}_{il}^{\text{H}}{{\mathbf{Q}}_{kl}}{{{\mathbf{\bar h}}}_{il}} \!+\! I_{il}^2\left[ {m \!-\! s} \right]{\left| {{\mathbf{\bar h}}_{kl}^{\text{H}}{{{\mathbf{\bar h}}}_{il}}} \right|^2} .
\end{align}
Submitting \eqref{DS} and \eqref{UI1} into \eqref{UI}, we can derive
\begin{align}
  \mathbb{E}\left\{ {{{\left| {{\text{U}}{{\text{I}}_{ki}}\left[ {m - s} \right]} \right|}^2}} \right\} = {p_i}{\left( {\sum\limits_{l = 1}^L {a_{kl}^*\left[ s \right]\left( {{I_{il}}\left[ {m - s} \right]{\mathbf{\bar h}}_{kl}^{\text{H}}{{{\mathbf{\bar h}}}_{il}}} \right)} } \right)^2} + {p_i}\sum\limits_{l = 1}^L {{{\left| {a_{kl}^*\left[ s \right]} \right|}^2}}  \notag \\
   \times \left( {I_D^2\left[ {m - s} \right]{\text{tr}}\left( {{{\mathbf{R}}_{il}}{{\mathbf{Q}}_{kl}}} \right) + I_D^2\left[ {m - s} \right]{\mathbf{\bar h}}_{kl}^{\text{H}}{{\mathbf{R}}_{il}}{{{\mathbf{\bar h}}}_{kl}} + I_{il}^2\left[ {m - s} \right]{\mathbf{\bar h}}_{il}^{\text{H}}{{\mathbf{Q}}_{kl}}{{{\mathbf{\bar h}}}_{il}}} \right) .
\end{align}
to finish the proof.
\end{appendices}

\bibliographystyle{IEEEtran}
\bibliography{IEEEabrv,Ref}

\end{document}